\documentclass[preprint,5p,times]{elsarticle}

\usepackage[T1]{fontenc}
\usepackage{amsmath}
\usepackage{amsthm}
\usepackage{bm}
\usepackage{breakurl}
\usepackage{breqn}
\usepackage[breaklinks]{hyperref}
\usepackage{amssymb}
\usepackage{listings}
\usepackage{gensymb}
\usepackage{fancyhdr}
\usepackage{url}
\usepackage{lineno}
\usepackage{xcolor,colortbl}
\usepackage{multirow}
\usepackage{algorithm}
\usepackage{algpseudocode}
\usepackage{mathtools}

\DeclarePairedDelimiter\ceil{\lceil}{\rceil}
\DeclarePairedDelimiter\floor{\lfloor}{\rfloor}

\algnewcommand{\LeftComment}[1]{\Statex \(\triangleright\) #1}

\definecolor{temporal}{RGB}{112, 186, 164}
\definecolor{red}{RGB}{219, 191, 186}
\definecolor{orange}{RGB}{191, 167, 214}
\definecolor{yellow}{RGB}{240, 230, 140}

\journal{Future Generation Computer Systems}

\begin{document}
\begin{frontmatter}

\title{Accelerating Range Minimum Queries with Ray Tracing Cores}

\author[a]{Enzo Meneses}
\author[a]{Crist\'obal A. Navarro\corref{author}}
\author[a]{H\'ector Ferrada}
\author[a]{Felipe A. Quezada}
\cortext[author] {Corresponding author.\\\textit{E-mail address:} cristobal.navarro@uach.cl}
\address[a]{Instituto de Informática, Universidad Austral de Chile.}

\begin{abstract}
During the last decade GPU technology has shifted from pure general purpose computation to the inclusion of application specific integrated circuits (ASICs), such as Tensor Cores and Ray Tracing (RT) cores. Although these special purpose GPU cores were designed to further accelerate specific fields such as AI and real-time rendering, recent research has managed to exploit them to further accelerate other tasks that typically used regular GPU computing. 
In this work we present RTXRMQ, a new approach that can compute range minimum queries (RMQs) with RT cores. The main contribution is the proposal of a geometric solution for RMQ, where elements become triangles that are placed and shaped according to the element's value and position in the array, respectively, 
such that the closest hit of a ray launched from a point given by the query parameters corresponds to the result of that query. Experimental results show that RTXRMQ is currently best suited for small query ranges relative to the problem size, achieving up to $5\times$ and $2.3\times$ of speedup over state of the art CPU (HRMQ) and GPU (LCA) approaches, respectively. Although for medium and large query ranges RTXRMQ is currently surpassed by LCA, it is still competitive by being $2.5\times$ and $4\times$ faster than HRMQ which is a highly parallel CPU approach. Furthermore, performance scaling experiments across the latest RTX GPU architectures show that if the current RT scaling trend continues, then RTXRMQ's performance would scale at a higher rate than HRMQ and LCA, making the approach even more relevant for future high performance applications that employ batches of RMQs.
\end{abstract}


\end{frontmatter}

\section{Introduction}
The graphics processing unit (GPU) has become a key resource for accelerating applications in science, technology and entertainment \cite{navarro2014survey, nickolls2010gpu, owens2008gpu} thanks to its high performance parallel compute architecture. In general, the GPU programming model is a general purpose abstraction that enables the development of scalable parallel algorithms to accelerate the solution of data-parallel problems. In the case of CUDA (Nvidia's GPU programming platform) there are four constructs that play an important role in this programming model; the \textbf{kernel} - which is the parallel code written by the programmer to be executed in GPU - and the \textbf{thread}, \textbf{block}, \textbf{grid} resource hierarchy \cite{nickolls2008scalable}. Each thread executes the kernel and they are spatially organized into blocks (1D, 2D, or 3D). Similarly, blocks are again spatially organized to form the grid (1D, 2D or 3D) which holds the entire description of resources to be used. Developing a GPU program  requires writing a kernel, where the parallel algorithm is, and specifying an amount of resources through the thread, block, grid hierarchy which is passed along with the kernel when being executed.

During the first decade of GPU computing (from $\sim$2006 to $\sim$2016), GPUs evolved mainly by increasing their number of floating point/integer units in the GPU chip, as well as by increasing the size and flexibility of the L1/L2 caches. Many of those improvements were highly transparent for the programmer, that is the applications would obtain extra acceleration just by switching to the new GPU hardware and recompiling the code. Although this trend continues in the present, by itself it is not sufficient to fulfill the needs of certain demanding applications, such as the training/inference of neural networks for artificial intelligence (AI), or the generation of photorealistic computer graphics. For those reasons, in the year 2017 Nvidia introduced tensor cores in their GPU chips, which are application specific integrated circuits (ASICs) devoted to accelerate artificial intelligence applications (AI), more specifically the matrix multiplications involved in the training/inference of neural networks. One year later, in 2018 Nvidia introduced the \textbf{ray tracing cores} (\textbf{RT cores}), a second ASIC devoted to accelerate the ray-triangle intersection test involved when launching rays into a scene of geometry, which is the most costly operation in the process of generating photorealistic real-time graphics.

\begin{figure*}[ht!]
    \centering
    \includegraphics[scale=0.45]{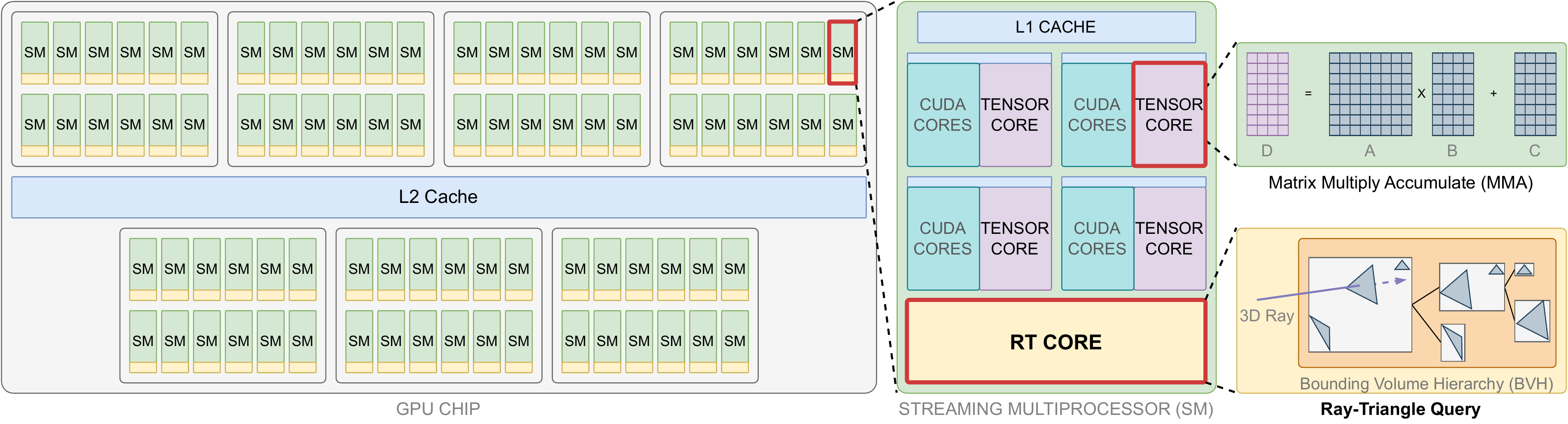}
    \caption{The modern GPU exhibits a hierarchical design. At the top level (left) there are large number of clusters of resources which are known as streamming multiprocessors (SMs). Inside each SM there are CUDA cores, Tensor Cores and Ray Tracing (RT) Cores. Tensor cores offer a hardware level matrix multiply accumulate (MMA) operation while the RT Core offers a fast ray-triangle query through a hardware implemented Bounding Volume Hierarchy (VBH) data structure.}
    \label{fig:gpu-arch}
\end{figure*}

As of 2023, GPUs offer a diverse ecosystem of resources as shown in Figure \ref{fig:gpu-arch}; from general purpose cores (\textit{i.e.}, floating point/integer units) to application specific ones. Although tensor and RT Cores were originally designed to accelerate specific tasks such as AI and graphics, respectively, recent advances in GPU Computing research have shown that other GPU applications can also benefit from using these ASICs. 
In the case of tensor cores, a significant amount of recent research has shown that they can indeed accelerate a diverse number of non-AI patterns such as summation \cite{carrasco2018analyzing,navarro2020gpu}, prefix sum \cite{dakkak2019accelerating}, fractal mapping \cite{NAVARRO2020158,QUEZADA202210}, FFT \cite{sorna2018optimizing, durrani2021fft, li2021tcfft}, stencil computations \cite{liu2022toward}, among others. In the case of Ray Tracing cores, the research is more recent and there are already successful adaptations for nearest neighbors search \cite{zhu2022rtnn, zhao2023leveraging}, generation of force directed graphs \cite{zellmann2020accelerating} and point location in meshes \cite{morrical2020accelerating, wald2019rtx}, reporting significant speedup over a regular GPU implementation. 
What makes GPU Ray Tracing cores an attractive tool for research is the fact that they offer a hardware level search mechanism based on the interaction of rays and triangles in 3D space. If a regular GPU application can be adapted to the RT pipeline, then it can obtain a significant acceleration over its regular GPU version. The research challenge is to find a proper spatial representation of the input data, and to design ray-triangle queries such that they serve as a search function. One well known computational problem that uses a high number of search operations is the Range Minimum Query (RMQ) \cite{fischer2011space,fischer2006theoretical}; 
it plays a relevant role in string processing applications such as document retrieval \cite{muthukrishnan2002efficient}, web search engines \cite{croft2010search,kobayashi2000information} and computational biology \cite{abouelhoda2004chainer}, among others, as its solution allows answering other queries such as the Lowest Common Ancestor (LCA) \cite{bender2000lca}. The possibility of accelerating RMQ with GPU RT cores constitutes the main motivation of this work, as it can benefit the aforementioned applications as well as many more that rely on RMQs.

This work presents a new method, named RTXRMQ, for solving batches of Range Minimum Queries (RMQs) in parallel using GPU RT Cores. The main contributions are:
\begin{itemize}
    \item The reformulation of the RMQ problem in the ray tracing model, such that the intersections between rays and triangles can now answer RMQ queries.
    \item A detailed description of the RTXRMQ implementation using OptiX \cite{parker2010optix}, as well as the optimizations.
    \item An experimental evaluation of performance, scaling and energy efficiency that is compared to the state of the art RMQ methods in CPU (\cite{Ferrada2016ImprovedRM}) and GPU \cite{polak2021euler}.
    \item The source code\footnote{RTXRMQ available at \url{https://github.com/temporal-hpc/rtxrmq}} is open and available to the community.
\end{itemize}
 Experimental evaluation shows that for small query ranges the proposed RTXRMQ is up to $5\times$ and $2.3\times$ faster than state of the art multi-core CPU \cite{Ferrada2016ImprovedRM} and GPU \cite{polak2021euler} methods, respectively, as well as competitive in energy efficiency. Also, its performance scales when switching up through the different generations of RT cores (Turing, Ampere, Lovelace); this makes RTXRMQ capable of scaling its performance for future GPUs at a higher rate than other approaches basd on regular GPU computation.

The rest of the manuscript is organized as follows: Section \ref{sec:rmq} presents the problem statement for RMQ, Section \ref{sec:rtcore-overview} presents an overview of the RT Core programming pipeline, Section \ref{sec:related-work} presents related work, Section \ref{sec:rmqs-rtcores} the new proposed RTXRMQ, Section \ref{sec:experimental-evaluation} the experimental evaluation and Section \ref{sec:conclusions} the discussion and conclusions of the work.

\section{Problem Statement: Range Minimum Query (RMQ)}
\label{sec:rmq}
The Range Minimum Query (RMQ) problem is defined as follows: given an array $X = [x_0, x_1, \cdots, x_{n-1}]$ of $n$ elements from a sortable set, and a pair of positions $l \leq r < n$, RMQ returns the position of the minimum value in that given range. It can be formally expressed as 
\begin{equation}
\text{RMQ}(l,r) = \text{argmin}_{l\leq k\leq r} x_k
\end{equation}

If the minimum exists more than once, then a common approach is to prefer the leftmost one. As an example, given $X = [9,2,7,8,4,1,3]$ and a range $[2,6]$, then $\text{RMQ}(2,6) = 5$, that is $x_5 = 1$.

RMQ was first studied in the context of a Compact Data Structure (CDS) \cite{Navarro2016,NM07}. In a theoretical work, Fisher and Heum showed in 2011 that RMQ can be solved in constant time using $2n+o(n)$ bits \cite{fischer2011space} when approached with a Depth-first Unary Degree Sequence (DFUDS) representation of Cartesian Trees \cite{BDMRRR05, Vu80}. In 2017, Ferrada and Navarro \cite{FN17} proposed an improved RMQ scheme that is faster and achieves $\sim 2.1n$ bits of space. One of its key aspects is the representation of the Cartesian Tree using Balanced Parentheses. Up to date, the solution of Ferrada and Navarro is considered one of the fastest CPU-based RMQ implementations and a reference as its source code has been made available to the community\footnote{Ferrada's library is available at \url{https://github.com/hferrada/rmq}}. The present work considers this library as well as a reference comparison in the experimental evaluation of Section \ref{sec:experimental-evaluation}. 

The RMQ problem is also closely related to the problem of determining the Lowest Common Ancestor (LCA) in ordinal trees \cite{AHU73}.
In fact, there is a linear reduction in both directions between these problems, RMQ and LCA, whose time is linearly proportional to the size of the input $n$.

Even if no compressed data structured is employed, RMQ is still an attractive research problem from the point of view of performance, as there are several applications such as computational genomics \cite{abouelhoda2004chainer}, web search engines \cite{croft2010search,kobayashi2000information}, document retrieval \cite{muthukrishnan2002efficient}, among others that would benefit from having an accelerated RMQ approach even if memory usage is sacrificed to a certain degree. Moreover, considering the recent advances in GPU hardware towards more parallelism and the inclusion of ASICs, there exists the opportunity to accelerate batches of RMQs in parallel with RT Cores. 


\section{Overview of RT Core Programming with OptiX}
\label{sec:rtcore-overview}
The RT Core programming pipeline was designed with real-time computer graphics in mind as the target application, although technically it is not limited to it. This pipeline is intended to provide a comfortable level of abstraction for programming hardware-accelerated lighting algorithms such as ray tracing.

Ray tracing is a computer graphics technique that simulates the behavior of light as it travels and interacts with various objects and surfaces in a scene, resulting in the generation of photorealistic digital images that accurately reflect the physical world. This process requires a significant amount of computational work and resources, therefore numerous efforts have been made to enhance the efficiency of this technique, particularly through the implementation of an acceleration structure (AS) called Bounding Volume Hierarchy (BVH), which has been noteworthy in its success. 

\begin{figure}[ht!]
    \centering
    \includegraphics[scale=0.56]{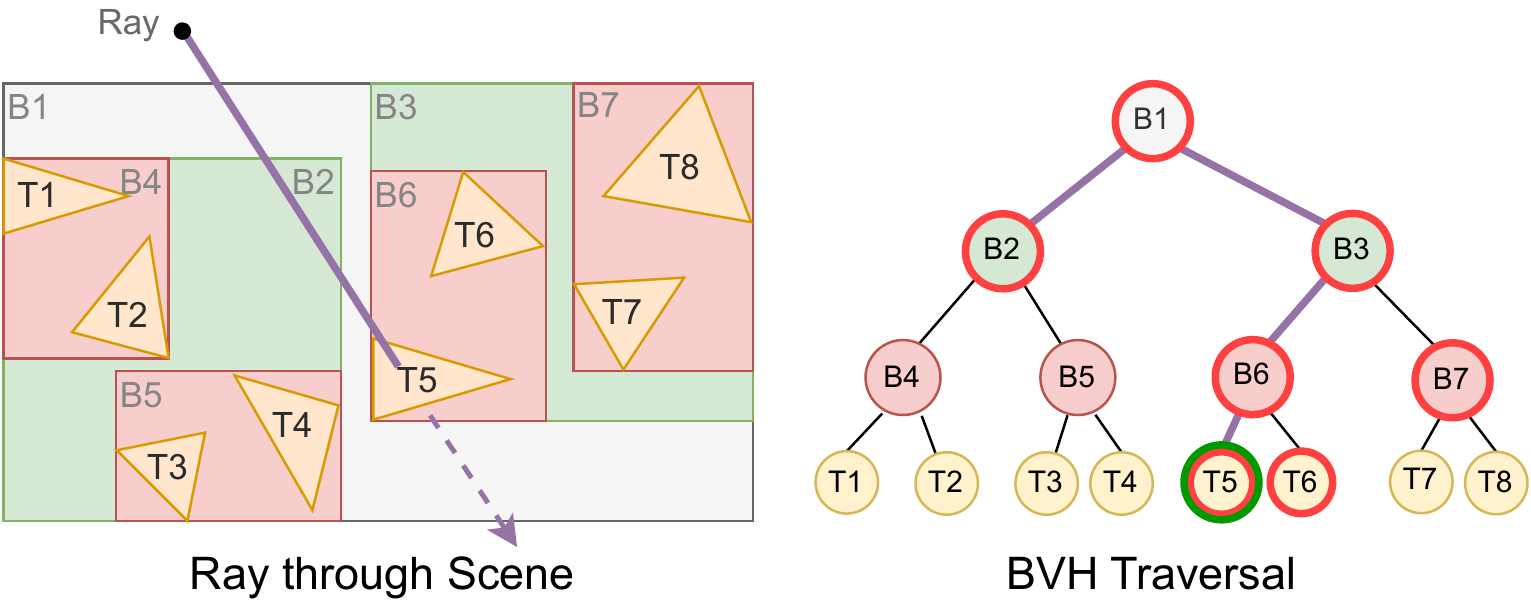}
    \caption{Example of how a ray traverses a BVH for a set of triangles. On the left, a view of how the BVH builds a hierarchy of bounding boxes (white, green and red boxes) for the triangles of the scene. On the right, the tree view of the BVH traversal, where only the nodes marked in red are required to visit in order to detect the collision between the ray and $T_5$, avoiding the need to access the entire set of triangles. Image inspired from Nvidia's developer blog \cite{NvidiaDeveloperBlog}.}
    \label{fig:bvh}
\end{figure}

The BVH is a tree-like data structure \cite{wald2007fast,klosowski1998efficient} that contains multiple bounding boxes of the scene arranged in a hierarchical manner. The BVH acts as a pathway for the light ray to traverse the scene, reducing the number of collision checks and improving efficiency. Figure \ref{fig:bvh} presents a visual intuition for the BVH.

In the process of ray tracing, rays are cast from a chosen position into the scene and traverse the BVH until they intersect with a primitive geometric shape, such as a sphere, square, triangle, or other similar object. Due to its intensive computational requirements, ray tracing techniques were primarily delegated to offline rendering, where all the calculations could be done in advance. As a result, real-time applications typically relied on raster-based rendering methods. In an effort to enable real-time performance for ray tracing, in 2018 NVIDIA introduced the Ray Tracing Core (RT Core) in their Turing architecture GPUs released in 2018. RT Cores accelerate BVH ray traversal as well as ray/triangle intersection testing using a hardware implementation of a custom BVH. These RT cores sit next to the CUDA and Tensor Cores of modern GPU chips.
As of 2023, NVIDIA is already in its third generation RT cores in the Lovelace architecture, reporting up to $\sim40\times$ of ray tracing performance improvement over a traditional GPU implementation (Turing provided the first factor of $10\times$ \cite{nvidiaTuring}, and Ada provides an extra factor of $4\times$ over Turing \cite{nvidiaAda}), making it a compelling tool to extend its performance improvement to other applications.


NVIDIA leverages a series of APIs such as NVIDIA Optix, Microsoft DXR and Vulkan to accelerate the ray tracing pipeline using RT Cores. The following explanations are in the context of OptiX as it is the API used for this work. 

\begin{figure}[ht!]
    \centering
    \includegraphics[scale=0.6]{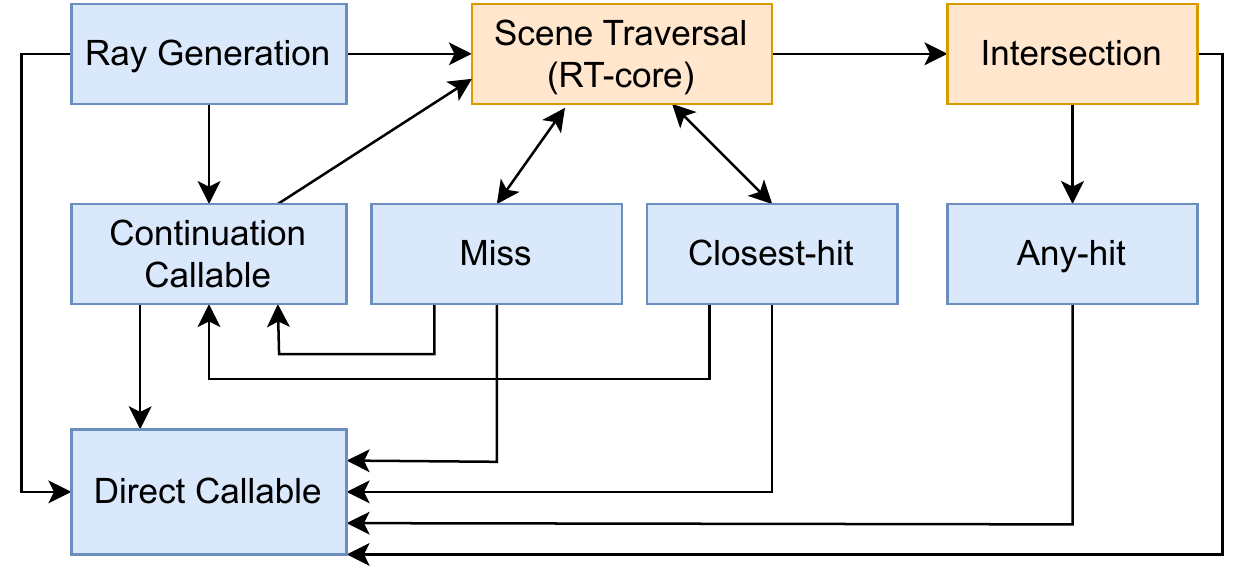}
    \caption{Optix pipeline execution graph. Arrows represent the execution direction. Blue stages are user-programmable, while oranges are handled internally via hardware implementation.}
    \label{fig:optix-pipeline}
\end{figure}

The ray launch pipeline of Optix is composed of 8 programmable stages depicted in Figure \ref{fig:optix-pipeline}.
The stages in blue represents user-programmable shaders, while orange are hardware implementations, executed via the RT Core.
The \textbf{Ray Generation} shader is executed at the start of the pipeline when a ray is generated from a specific 3D location in space and launched through the scene. A user-defined payload can be attached to the ray, which can be accessed in various stages of the pipeline. \textbf{Scene Traversal} and \textbf{Intersections} are efficiently managed by OptiX and the RT Cores. During these stages, the ray passes through the BVH and undergoes triangle intersection testing. \textbf{Miss} and \textbf{Closest-hit} shaders, which can be programmed, are triggered for each ray at the end of the traversal either in the absence of a collision or with the closest collision detected, respectively. The \textbf{Any-hit} is called in a per collision basis and can also be programmed. Finally the \textbf{Direct Callable} and \textbf{Continuation Callable} are non-predefined programs used to finalize calculations or continue with new rays, respectively. The Any-Hit and Closest-Hit stages can be disabled from the pipeline in order to increase performance.

A typical OptiX program begins with a scene comprised of geometric objects, typically formed by triangles. The triangles of the scene are passed to OptiX to build the Acceleration Structure (AS), which can further be optimized into an Instance Acceleration Structure (IAS) by organizing ASs hierarchically. Next, the various programs of the pipeline are created together with the Shader Binding Table (SBT), which binds the programs with objects in the scene. Finally, the program can be launched with a grid of rays, analogous as to how a CUDA kernel is launched along with a grid of threads.
It is worth mentioning that OptiX can interoperate with CUDA and share device memory pointers in order to do traditional GPU computation on the same data processed by RT cores. Such feature is relevant to CUDA developers as they can now consider RT cores to further accelerate some of the processing stages of their CUDA implementation.

\section{Related Work}
\label{sec:related-work}
Related work can be grouped into two categories; i) parallel RMQs on GPU and ii) other uses of RT Cores.

\subsection{Parallel RMQs on GPU}
Works on GPU-based parallel RMQs do exist and use the traditional GPU programming model.
Soman \textit{et al.} proposed in 2010 and 2012 a discrete range search primitive for GPU \cite{soman2010efficient,soman2012discrete}, which can be used to answer batches of RMQs through the LCA scheme up to $6\times \sim 8\times$ faster than a multi-core CPU solution. In their work, the authors implemented a succinct tree representation of the Cartesian tree that is GPU efficient, and their method uses a three level structure, where the first level minimizes memory usage, level 2 focuses on fast querying through batches and level 3 focuses on using minimal additional space. 

In 2021, Polak \textit{et al.} proposed an efficient way for doing Euler tours on GPU \cite{polak2021euler} and implemented a GPU version of the Schieber and Vishkin Inline LCA algorithm \cite{schieber1988finding} which solves batches of LCA queries using Euler tours. The authors report speedups between $22\times \sim 52\times$ over a sequential CPU implementation of the same Inline algorithm, and have made their source code available to the community. Considering that LCA can be used to answer RMQ queries, the work of Polak \textit{et al.} can be considered one of the fastest GPU-based RMQ solvers that is also available. The experimental evaluation of Section \ref{sec:experimental-evaluation} includes this implementation for comparisons.

\subsection{Other uses of RT cores}
Research on leveraging RT cores dates back to 2019 (one year after the introduction of RTX GPUs in 2018) for accelerating point location in meshes. In 2019, Salmon and McIntosh-Smith proposed a ray tracing approach to accelerate the Monte Carlo particle transport algorithm \cite{salmon2019exploiting}, where RT cores were used to know if a point is inside or outside a mesh. This was done by building the RT core bounding volume hierarchy (BVH) for the mesh and launching one ray per point. For each ray, if the number of intersections are odd then the point is inside the mesh, otherwise it is outside (even). The authors report a speedup of up to $20\times$ over a non-RT GPU approach. Also in 2019, Wald \textit{et al.} proposed a way to accelerate point location in unstructured tetrahedral meshes \cite{Wald2019RTXBR}. In their work, the authors propose three variants that leverage RT cores, achieving a speedup of $4\times$ over an already efficient CUDA based BVH implementation.    

Until now, one of the main uses of RT cores has been the computation of nearest neighbors, which has produced very successful research cases. In 2020, Zellmann \textit{et al.} \cite{zellmann2020accelerating} proposed a method for computing 2D fixed radius nearest neighbors (FRNN) using RT cores in the context of accelerating force-directed graph drawing, achieving speedups from $4\times$ up to $13\times$ over a regular CUDA implementation. The approach is based on creating axis aligned bounding boxes (AABBs) of the size of the radius around each point followed by launching infinitesimal rays, one for each point again, which allows registering the nearest neighbors for all points. Such approach works because when a ray intersects an AABB, it triggers an intersection function which can be programmed to know if the point from the intersecting ray is within the radius circle of the nearest neighbor. By registering all intersections in the same way (except with itself), one computes the FRNNs of all points. This technique is also known as the inverse radius search. 
In 2021, Evangelou \textit{et al.} \cite{Evangelou2021RadiusSearch} also used an inverse radius search approach to compute both the k-nearest neighbors (kNN) and FRNN in 3D with RT cores. The authors report favorable speedup in almost all tests, and in the best cases speedups of up to $30\times$ over FLANN, a known kd-tree GPU implementation. In 2022, Zhu proposed additional optimizations for 2D/3D kNN and FRNN with RT cores \cite{zhu2022rtnn}; query scheduling and query partitioning, reaching speedups of up to $65\times$. Recently, in 2023 Zhao \textit{et al.} managed to accelerate the nearest neighbors search phase of particle-based simulations using RT cores \cite{zhao2023leveraging}, reporting speedups of up to $1.6\times$ over grid based solutions on GPU which are known to be already efficient.

Given the current scenario of GPU-based RMQ solutions and what uses have been done with RT cores, the idea of leveraging RT cores to solve batches of RMQs becomes a research opportunity that can significantly contribute in the knowledge on what new patterns can be accelerated with RT cores.

\section{RTXRQM: Computing RMQs with RT Cores}
\label{sec:rmqs-rtcores}
The key idea for processing RMQs with RT cores is to take advantage of the fact that when placing the elements of any unordered array $X = [x_0, x_1, \cdots, x_{n-1}]$ along an axis by their value, they become automatically ordered, similar in spirit to how the counting sort algorithm works. From there, the challenge is to design a ray/triangle intersection scheme that can answer $\text{RMQ}(i,j)$ queries efficiently. In order to better understand how this intersection scheme works, we first illustrate the simpler case of computing the global minimum of an array with RT cores, which is also the case of $\text{RMQ}(0,n-1)$.

\subsection{Simpler case: Searching the Minimum with RT cores}
\begin{figure}[ht!]
\centering
\includegraphics[scale=0.1]{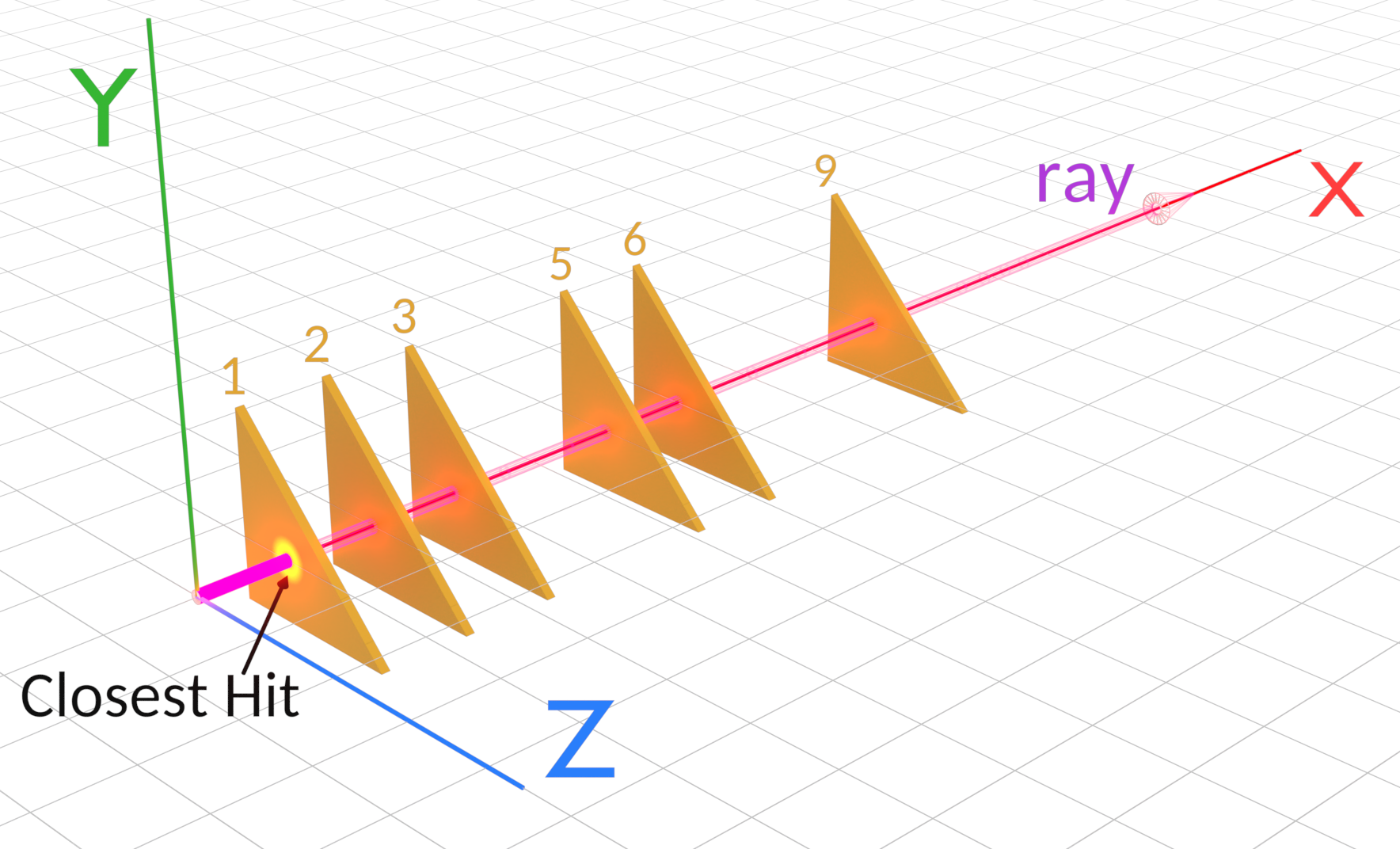}
\caption{Computing the minimum of $[5,3,1,9,6,2]$ with RT cores.} 
\label{fig:trivial_example}
\end{figure}

The RT core approach to compute the minimum begins by creating one geometric primitive (\textit{e.g.}, a triangle) for each element of the array and place it along the $X$ axis in the position of the element's value. Then, when launching a ray along that aixs from $-\infty$ (or from any position to the left of all elements) to $+\infty$, the closest hit event will be triggered with the object associated to the smallest element of the array, as shown in Figure \ref{fig:trivial_example}. Until now, the key aspect to highlight is that when placing objects in the position given by the array values, these objects become naturally ordered along the $X$ axis and the collision of rays can immediately answer minimum or maximum queries.
Solving RMQs with RT cores is an extension of this idea that not only places objects by the element's value, but also uses the elements position to define their shape. 

\subsection{Solving RMQs with RT Cores}
\label{subsec:rtxrmq-base}
Answering RMQs with RT cores shares the same design principle of computing the minimum with RT cores, \textit{i.e.}, to place objects in space by their value and have them ordered, but also considers the position of each value for the objects shapes and the query indices for the origin of the ray. These two differences are key design aspects that allow answering $RMQ(l,r)$ queries efficiently avoiding the collision with unnecessary objects outside the $[l,r]$ range.

\begin{figure}[ht!]
\centering
\includegraphics[scale=0.15]{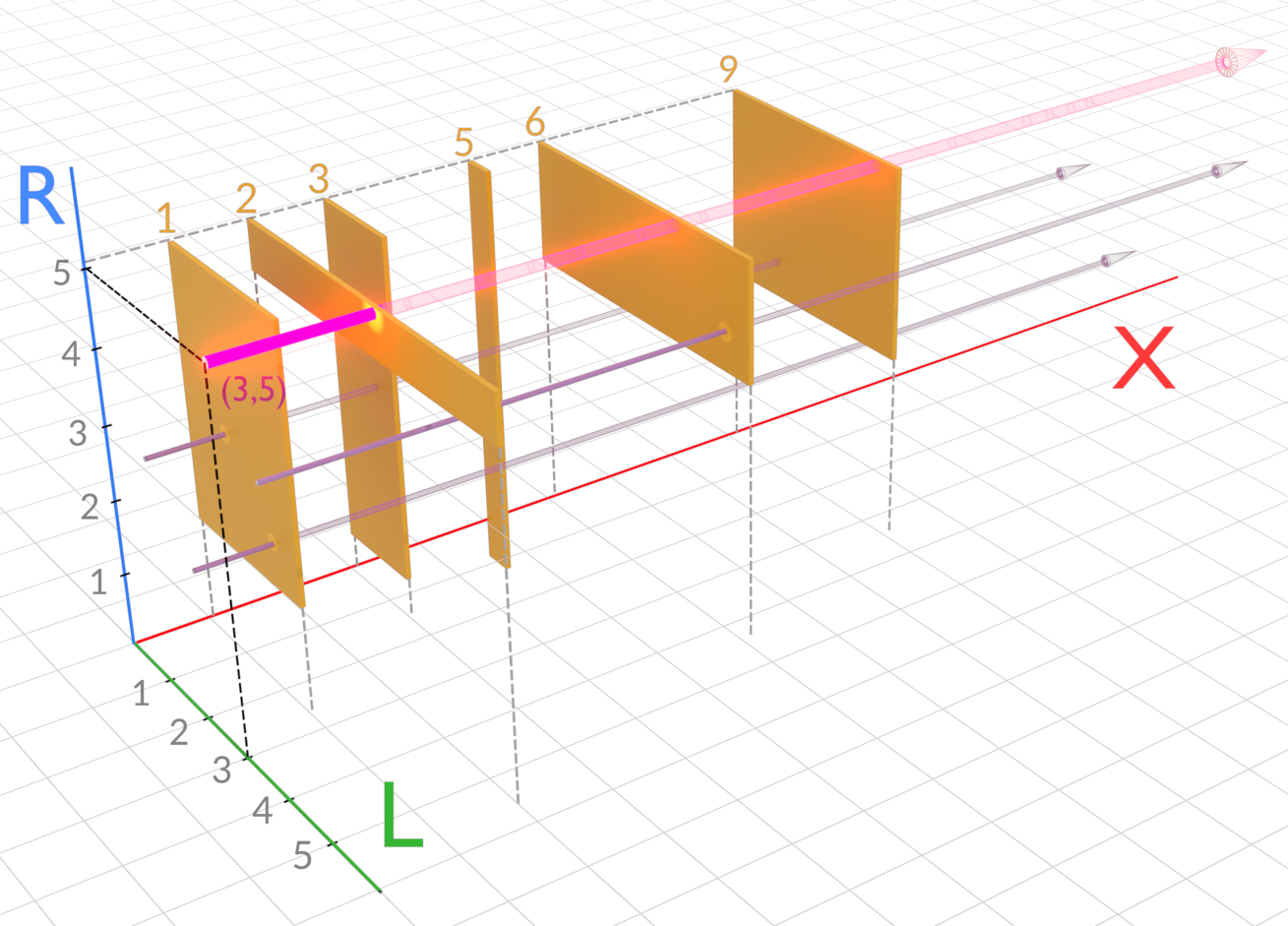}
\caption{Example of solving RMQ with RT cores. Geometry built from the array $[5, 3, 1, 9, 6, 2]$ and the query highlighted in magenta is $\text{RMQ}(3,5) = 2$. The magenta and purple rays constitute the RMQ batch.} 
\label{fig:geometry_example}
\end{figure}

Given that RT cores work in 3D space, and one dimension is already being used for the element values, the other two dimensions remain available and can be used to define the object shapes as well as how rays will be launched in the context of a $\text{RMQ}(l,r)$ query.  
The idea is to launch one ray per $\text{RMQ}(l,r)$ query such that it only intersects elements in the range $[l,r]$ of the input array. To achieve this, rays are launched from the coordinate $(-\infty, l, r)$ ($l,r$ of their corresponding query) with direction $(1,0,0)$ which is the dimension of the values. With this, the closest hit of that ray now answers the query $\text{RMQ}(l,r)$ as it hits the object with the minimum value in $[l,r]$. Moreover, given that there are multiple RT cores in a GPU, many rays (queries) can be processed in parallel for the same geometry built once. 
Figure \ref{fig:geometry_example} illustrates how a batch of RMQs are solved in parallel using the approach recently described. 

\begin{figure}[ht!]
\centering
\includegraphics[scale=0.5]{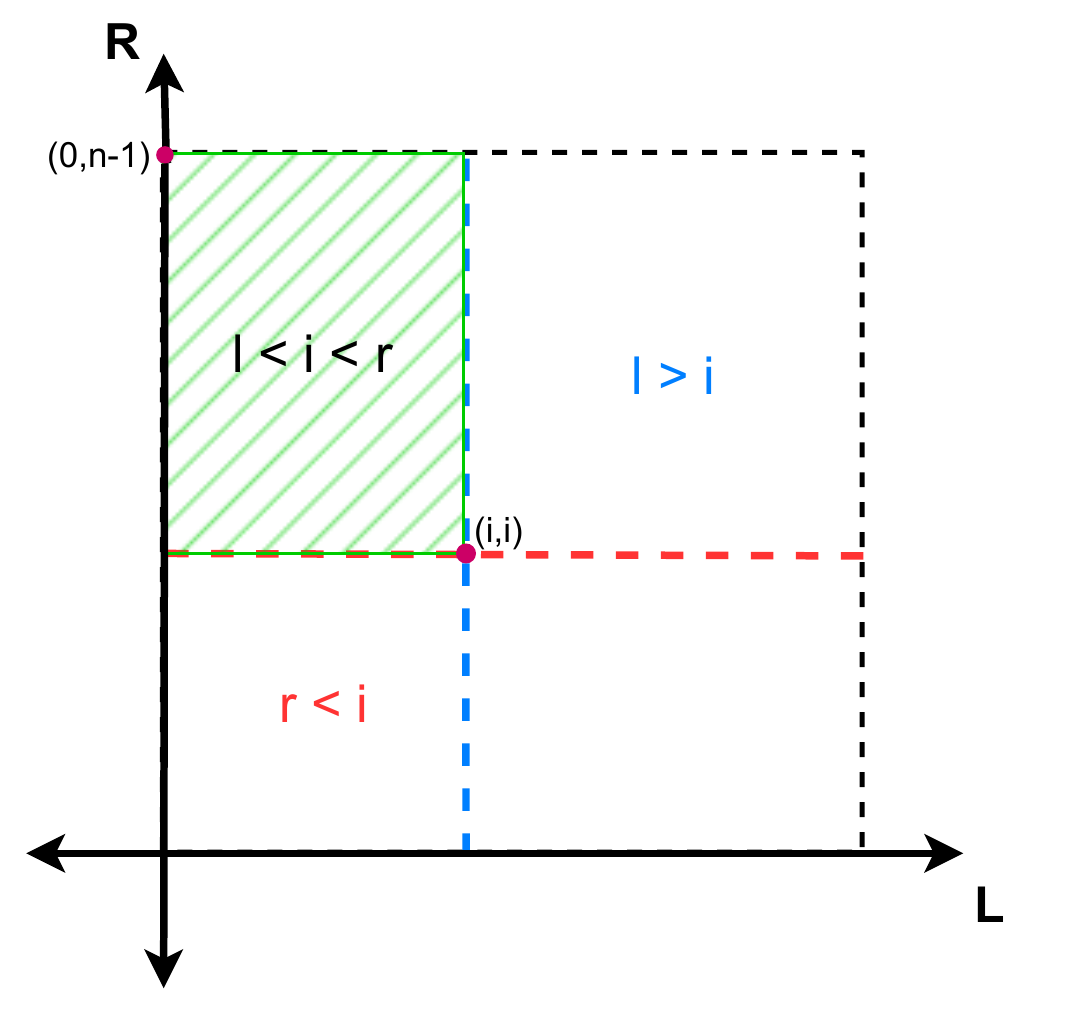}
\caption{View from the $L,R$ plane of the 3D scene. With this design, rays only intersect the $[l,r]$ range of a given query $\text{RMQ}(l,r)$.} 
\label{fig:geometry_construction_theory}
\end{figure}

In terms of the geometry, elements of the array theoretically correspond to rectangles which are located along the X axis, according to the element's value and perpendicular to the axis. In terms of its shape (width and height), it is defined in such a way that the rectangle for the element at position $i$ covers the biggest and smallest possible queries that contain the element, extending from $(0,n-1)$ to $(i,i)$. This means that for any element and $\text{RMQ}(l,r)$ query, if the lower limit of the query ($l$) is after the position of the element, then the query ray will pass on the right of the element's rectangle without intersecting. Similarly, if the end of the range ($r$) is before the element's position, then the ray will pass below its rectangle without intersecting again. Only if the element is in the range, \textit{i.e.}, $x_i \in X[l..r]$, then the ray will intersect its corresponding rectangle. Figure \ref{fig:geometry_construction_theory} illustrates a concept view of the scheme recently described, where a rectangle is represented by the green region, and rays will hit or pass depending on where they are launched from in the $L,R$ plane. 

\begin{figure}[ht!]
\centering
\includegraphics[scale=0.4]{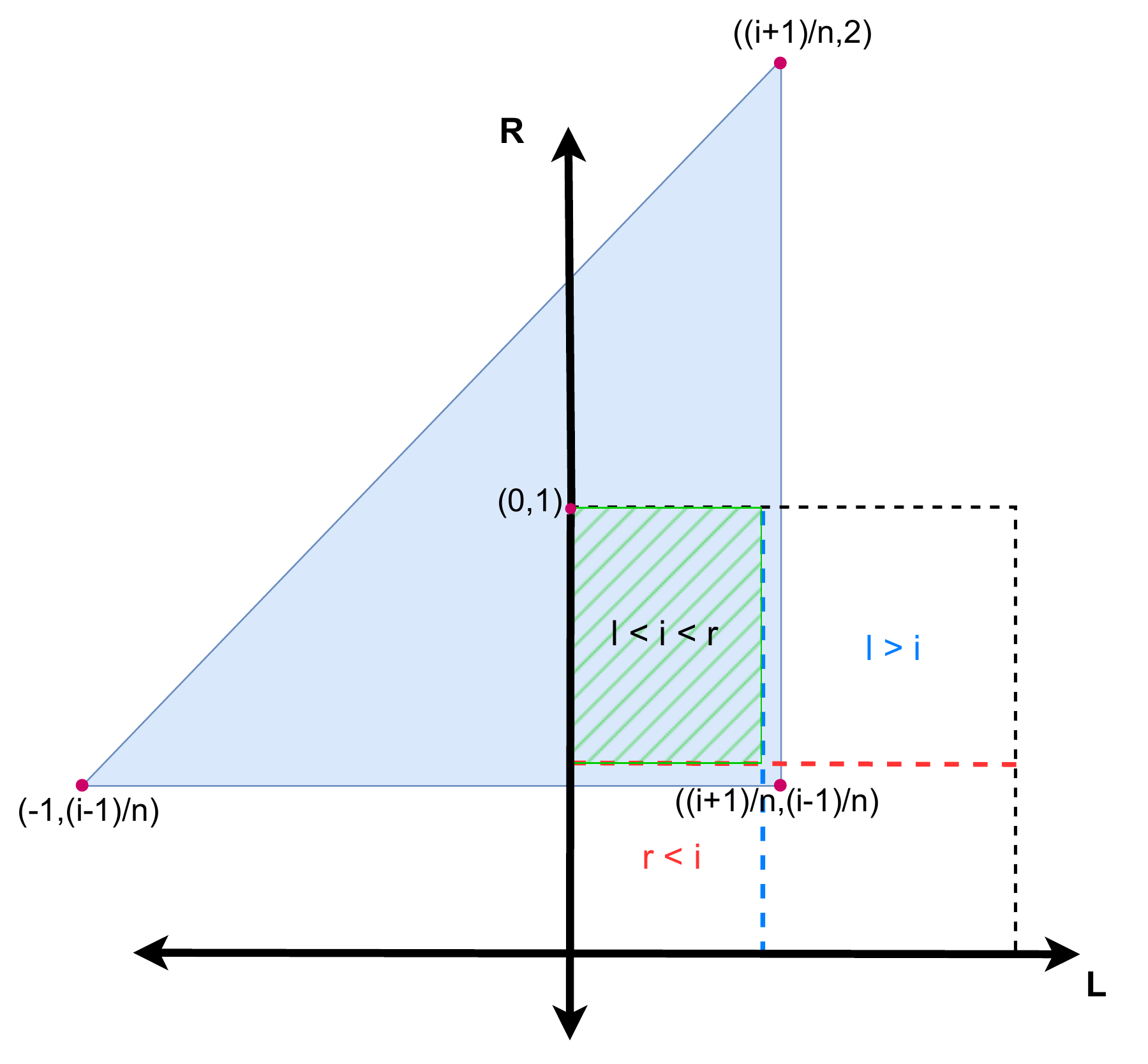}
\caption{Concept geometry for solving RMQ with RT cores and implemented geometry, viewed from the $L,R$ plane} 
\label{fig:geometry_construction}
\end{figure}

It is worth noticing that all possible queries are contained in the region $[0,n-1] \times [0,n-1]$ of the $L,R$ plane, which corresponds to the smallest and largest values of $l$ and $r$ in the $L$ and $R$ axes, respectively. In terms of GPU implementation, it is more efficient to build the geometry using triangles instead of rectangles, and use a normalized space as shown in Figure \ref{fig:geometry_construction}. 

The reason for using triangles is because the collision resolution between rays and triangles is accelerated by hardware (RT core), increasing performance, while for other primitives it does not. Figure \ref{fig:geometry_construction} shows how triangles are shaped to cover the corresponding rectangle. An extra border of one normalized unit is added to the right and bottom sides of each triangle in order to account for the rays that pass through the edge of a triangle, otherwise the RT core model does not detect it as an intersection. 
This need of adding borders is related to the fact that OptiX ray-triangle intersection is watertight for adjacent triangles, that is, a ray shot at the border of two adjacent triangles will only hit one of them. This means that with no adjacent triangle, a ray passing through the border may or may not hit the triangle depending on its orientation and the position of the border relative to the triangle. In the case of RTXRMQ rays passing through the bottom and right border are not considered as a hit, thus requiring the triangle to cover the ranges $[0,i+1)$ horizontally and $(i-1,n-1]$ vertically. Having a border thickness of one normalized unit allows ensuring the edges are properly covered, while not affecting the correctness as it is insufficient to catch the rays coming from other neighbor queries. It is also worth clarifying that the extra space introduced by the triangle's top and left vertices is not relevant for the process as it lies outside the query space $[0,n-1] \times [0,n-1]$. Regarding the normalization, it helps in numerical accuracy as well as in performance for the hardware implemented BVH. The implementation for this construction is shown in Algorithm \ref{alg:gen_triang}.
\begin{algorithm}[ht!]
\caption{Triangle Generation}
\label{alg:gen_triang}
\begin{algorithmic}
	\Require $\bm{A}$: Array, $\bm{n}$: size, $\bm{i}$: index
    \State$x \gets A_i$ \Comment{Array element}
    \State $l \gets (i+1)/n$
    \Comment{Compute normalized $l,r$ values}
    \State $r \gets (i-1)/n$
    \State $v_0 \gets (x, l, r)$
    \Comment{Generate the triangle vertices}
    \State $v_1 \gets (x, l, 2)$
    \State $v_2 \gets (x, -1, r)$
\end{algorithmic}
\end{algorithm}

Once the geometry is built, the geometric model is ready to answer any number of RMQ queries in parallel, only limited by the hardware memory. To answer a given set of $\text{RMQ}(l,r)$ queries, rays are launched from the $L,R$ plane towards the positive $X$ axis, from an $X$ position that is before any triangle, for example from $(-1,l,r)$ towards $(1,0,0)$. This is done in the ray generation shader as illustrated in Algorithm \ref{alg:raygen}.
\begin{algorithm}[ht!]
\caption{Ray Generation}
\label{alg:raygen}
\begin{algorithmic}
	\Require $\bm{G}$: Geometry, $\bm{l}$: left, $\bm{r}$: right, $\bm{n}$ size
    \State $\vec{r} \gets (\Theta, \frac{l}{n}, \frac{r}{n})$
    \Comment{Ray origin, $\Theta$ smaller than all elements}
    \State $\vec{d} \gets (1, 0, 0)$ \Comment{Ray direction}
    \State $p \gets 0$ \Comment{Payload}
    \State $t_{\min} \gets 0$
    \State $t_{\max} \gets \inf$
    \State $r_{\text{time}} \gets 0$ \Comment{Ray time}
    \State $\text{optixTrace}(G, \vec{r}, \vec{d}, t_\min, t_\max, r_{\text{time}}, p)$
    \State $\min \gets (p + \Theta)$ \Comment{Result of $\text{RMQ}(l,r)$}
    \State \Return $\min$
\end{algorithmic}
\end{algorithm}

All launched rays traverse the scene in parallel using the hardware-implemented BVH of the RT cores. Whenever a ray intersects a triangle, the Closest-hit shader is called. The implementation of this shader is shown in Algorithm \ref{alg:closest_hit}, where the $t$-value is stored in the payload and represents the distance traveled by the ray until the closest hit. Finally, for each ray, the minimum value for its $\text{RMQ}(l,r)$ query is computed using the origin and distance from the $t$-value.
\begin{algorithm}[ht!]
\caption{Closest-hit}
\label{alg:closest_hit}
\begin{algorithmic}
    \State $t \gets \text{optixGetRayTMax()}$
    \Comment{$t$-value of closest hit}
    \State $\text{optixSetPayload}(t)$
\end{algorithmic}
\end{algorithm}

OptiX API works with floating point numbers, thus building triangles from integer indices needs to be done with caution. For example, when using \texttt{FP32} floats, a simple cast from \texttt{int} to \texttt{float} allows a maximum array size of $2^{24}$ because of the $23+1$ bits of precision before running into wrong RMQ results. For larger arrays, a different transformation from int to float needs to be used, such as the one presented in Algorithm \ref{alg:int_to_float}. 
\begin{algorithm}[ht!]
\caption{Alternative Int to float Transform}
\label{alg:int_to_float}
\begin{algorithmic}
    \Require $x_{\text{int}}$ \Comment{$x \in \mathbb{Z}$}
    \State $E = \floor{x_{\text{int}} / 2^{23}}$ \Comment{Exponent}
    \State $M = x_{\text{int}} \mod 2^{23}$ \Comment{Mantissa}
    \State $q = (M + 2^{23}) / 2^{24}$ \Comment{$q \in \mathbb{R}$}
    \State $x_{\text{float}} = q \cdot 2^E$ 
    \State \Return $x_{\text{float}}$
\end{algorithmic}
\end{algorithm}

However, this solution does not scale well in performance because placing too many triangles one behind another from the ray's viewpoint produces too many bounding box intersections between the ray and the internal nodes of the BVH, specially when the closest hit is a triangle on the back of the geometry, with a complexity of $O(n \log n)$ in the worst case. For this reason, a block based extension is proposed to support large problem sizes.

\subsection{Block-Matrix Design for Large Inputs}
\label{subsec:rtxrmq-blocks}
\begin{figure}[ht!]
\centering
\includegraphics[scale=0.25]{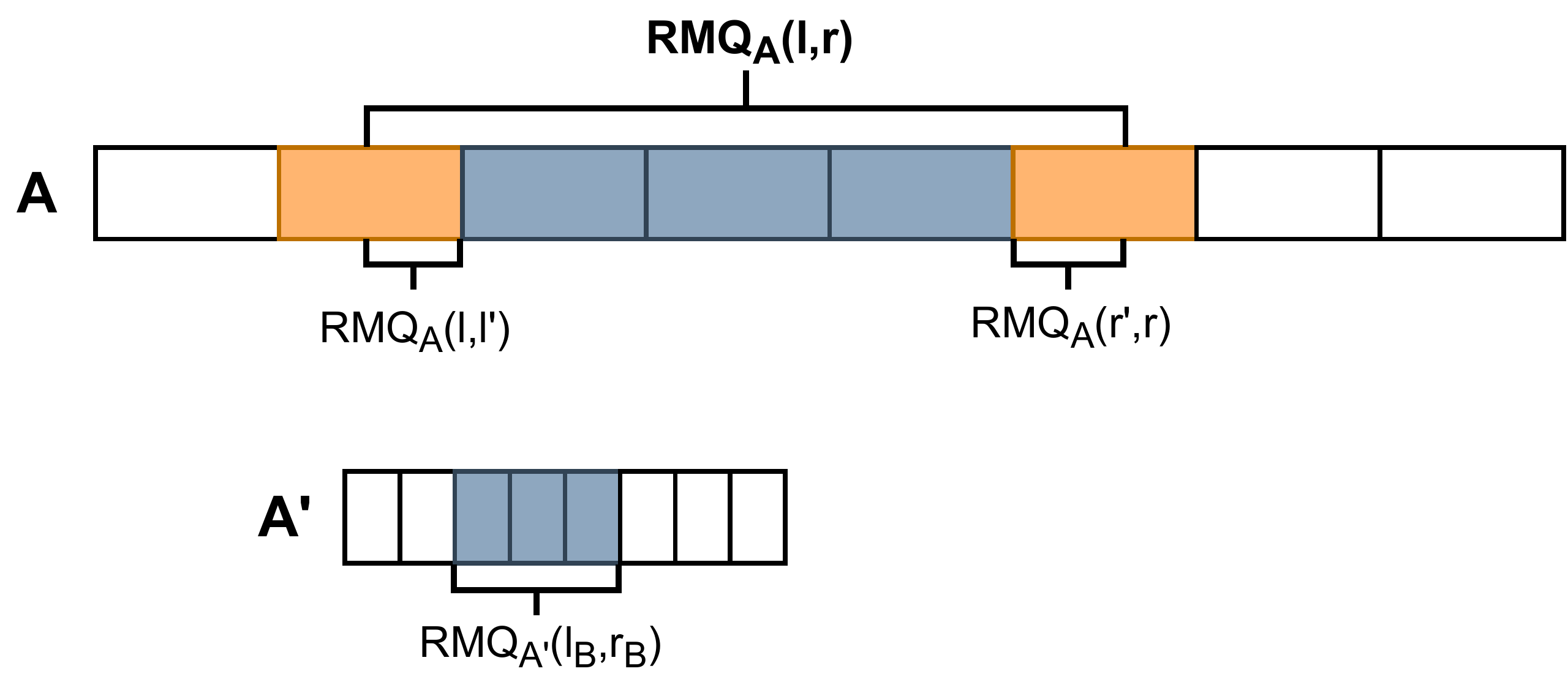}
\caption{$\text{RMQ_A}(l,r)$ can be expressed as the minimum between $\text{RMQ_A}(l,l')$, $\text{RMQ_A}(r',r)$ and $\text{RMQ_{A'}}(l_B,r_B)$ where $A'$ is an array for the minimums per block of $A$, $l_B$ and $r_B$ are the indices of the left-most and right-most blocks completely covered by the original query.} 
\label{fig:rmq_blocks}
\end{figure}

The proposed extension to support inputs of $n > 2^{24}$ is to subdivide the $\text{RMQ}(l,r)$ query into three smaller queries by partitioning the array into blocks and storing the minimums per block in another array. This array partitioning is just performed once for the input as a prepossessing stage, as illustrated in Figure \ref{fig:rmq_blocks} where $A$ is the original input array and $A'$ the block minimums one. 

\begin{figure}[ht!]
\centering
\includegraphics[scale=0.38]{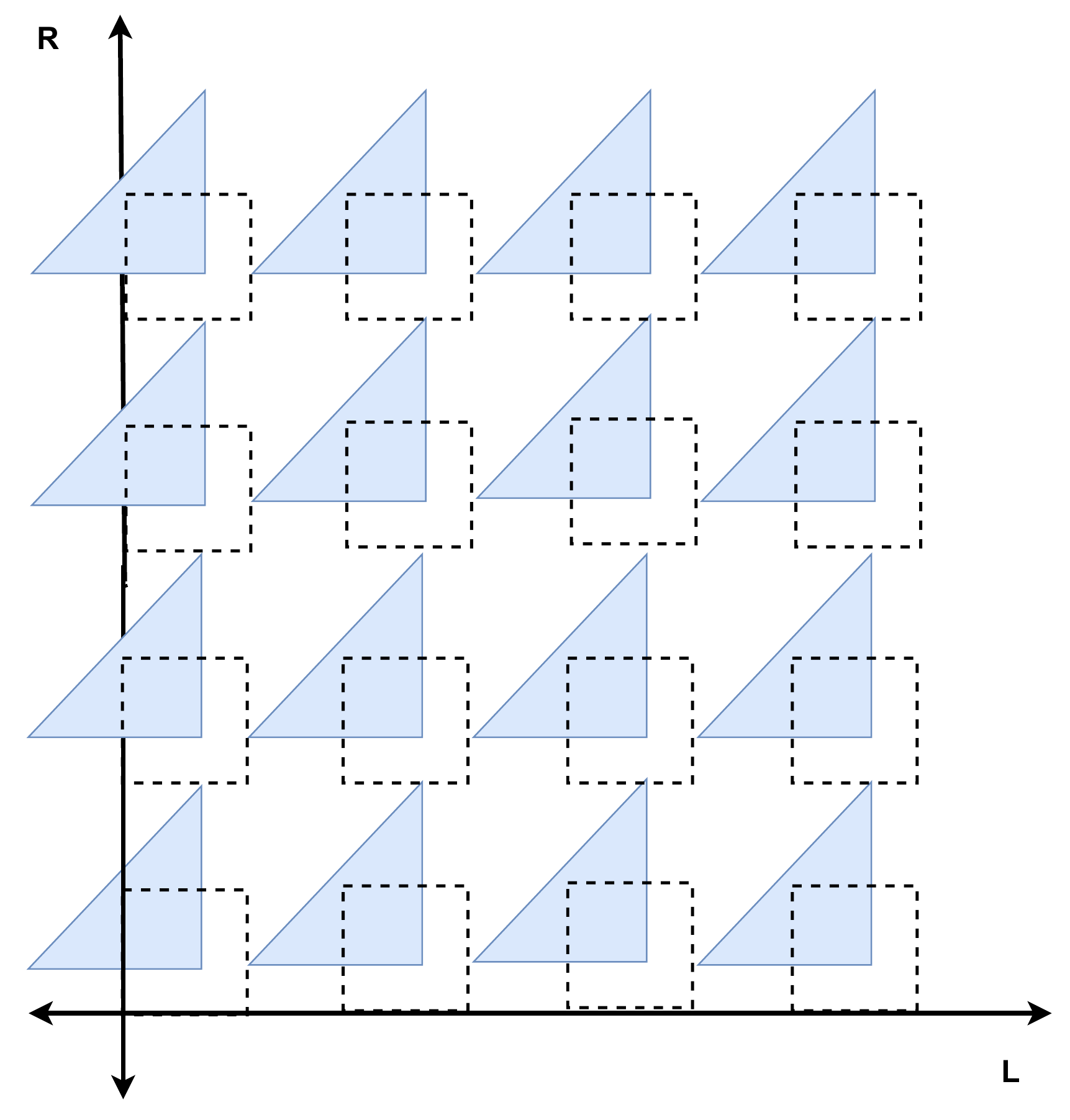}
\caption{Concept of geometry for solution dividing the array in blocks} 
\label{fig:geometry_blocks}
\end{figure}

Once the block minimums array is obtained, the idea is to first find the minimum of the block minimums just for the blocks that are completely covered by the query range, and then compare this temporal minimum to the minimums of the partially covered blocks which will require RT core processing. This RT core processing is done over a matrix of moderate-size geometry sets, one for each block, as shown in Figure \ref{fig:geometry_blocks}.   

The benefit of separating the domain into blocks is that it limits the number of triangles that a single ray can hit and also helps the BVH at doing a more efficient space partition, reducing the number of accesses to the internal nodes for each ray. Also, using a matrix-like layout of blocks instead of a linear one helps at keeping the sets closer to the origin where there is a more favorable floating point density and numerical accuracy.

The block-matrix triangle generation is presented in Algorithm \ref{alg:gen_triang_blocks}. 
\begin{algorithm}[ht!]
\caption{Block-matrix Triangle Generation}
\label{alg:gen_triang_blocks}
\begin{algorithmic}
	\Require $\bm{A}$: Array, $\bm{i}$: index, $\bm{n}$:size, $\bm{n}_{\bm{b}}$: block size
    \State $x \gets A_i$ \Comment{Array element}
    \State $i_b \gets i / n_b$ \Comment{Block index}
    \State $i_l \gets i \mod n_b$ \Comment{Local index}
    \State $B \gets \ceil{\frac{n}{n_b}}$ \Comment{Number of blocks}
    \State $b_x \gets (i_b + 1) \mod B$ 
    \Comment{Block-matrix $b_x,b_y$ coords}
    \State $b_y \gets (i_b + 1) / B$\Comment{$i_b=0$ reserved for block minimums}
    \State $l \gets (i_l + 1)/B + 2b_x$
    \Comment{Compute $l,r$}
    \State $r \gets (i_l - 1)/B + 2b_y$
    \State $v_0 \gets (x, l, r)$
    \Comment{Compute triangle vertices}
    \State $v_1 \gets (x, l, 2b_y+2)$
    \State $v_2 \gets (x, 2b_x-1, r)$
\end{algorithmic}
\end{algorithm}
The main difference with respect to Algorithm \ref{alg:gen_triang} is that it applies an $L,R$ offset to the triangles to place them at their corresponding block coordinate, and then each triangle is placed relative to this offset. 
Algorithm \ref{alg:raygen_blocks} shows how the ray generation shader works for the block-matrix approach. 
\begin{algorithm}[ht!]
\caption{Block-Matrix Ray Generation}
\label{alg:raygen_blocks}
\begin{algorithmic}
	\Require $\bm{G}$: geometry, $\bm{l}$, $\bm{r}$, $\bm{n}$, $\bm{n}_{\bm{b}}$: block size, $\bm{B}$: $\#$ blocks
    \State $b_l \gets l / n_b$ \Comment{left block index}
    \State $b_r \gets r / n_b$ \Comment{right block index}
    \LeftComment{Case \#1 query is in one block}
    \If {$b_l = b_r$} 
        \State RT core $\text{RMQ}(b_l^{\text{begin}}, b_l^{\text{end}})$  \Comment min stored in payload
        \State \Return 
    \EndIf

    \LeftComment{Case \#2 query involves two or more blocks}
    \State $r_1 \gets$ RT core $\text{RMQ}(l,b_l^{\text{end}})$ \Comment{Left partial block}
    \State $r_2 \gets$ RT core $\text{RMQ}(b_r^{\text{begin}},r)$ 
    \Comment{Right partial block}
    \State $r_3 \gets \textsc{max\_float}$
    \If {$b_r - b_l > 1$} \Comment{Minimum of full covered blocks}
        \State $r_3 \gets$ block-level RT core $\text{RMQ}(b_l+1,b_r-1)$
    \EndIf
    \State \Return $\min(r_1,r_2,r_3)$

\end{algorithmic}
\end{algorithm}

The algorithm begins by computing the left and right block indices. Then, the algorithm branches into two cases; Case \#1 the query is contained in one block (a special favorable case) thus only one RT core RMQ is needed, and Case \#2 the general case where the query is contained in two or more blocks. The general case computes from two to three RT core RMQs (three if there are full covered blocks) and returns the minimum of them.

As for the closest hit shader, Algorithm \ref{alg:closest_hit} can be reused as it is, or alternatively one can modify it to update the payload if the $t$-value is less than what is already stored in the same payload from previous rays; this last adaptation would replace the $\min(r_1,r_2,r_3)$ of Algorithm \ref{alg:raygen_blocks}.

Given that the floating point accuracy depends on the magnitude of the values (\textit{i.e.}, corresponds to the mantissa multiplied by the last bit ($2^{-23}$), higher numerical accuracy is obtained by placing objects closer to the origin. In this approach, for a given size array $n$ and block size $BS$, the needed precision is $1 / BS$ and the obtained precision is calculated from the furthest point from the origin in square coordinates (except those outside the search space of the rays). Therefore, to maintain correctness, all combinations of $n$ and $BS$ must satisfy the following inequality:
\begin{equation}
2^{\floor*{ \log_2 \left( 2 \ceil*{ \sqrt{n/BS}\ } \right) } } \times 2^{-23} \leq 1/BS.
\end{equation}

In general, smaller block sizes allow working with larger arrays while maintaining correctness. Currently, OptiX numerical precision limits require the block size to be less or equal than $2^{18}$, and the number of blocks less or equal than $2^{24}$. Additionally, currently OptiX has a limit of $2^{29}$ primitives (\textit{i.e.}, triangles) that a geometry acceleration structure can handle, and a limit of $2^{30}$ rays in a single launch.

Lastly, solving $\text{RMQ}(l,r)$ queries on the block minimums array can be solved in two ways. The first is to solve it again with the RT core proposal, by building another geometry and placing it in the scene with the geometries for each block. The second approach is to build a lookup table, \textit{i.e.}, a matrix with every possible combination of queries. The main differences are that the lookup table needs as much memory as the square of the number of blocks, but compensates by replacing the block level ray launches with constant time reads. Preliminary experimentation showed that building another acceleration structure resulted in faster performance than the lookup table. Therefore, in the following Sections, RTXRMQ will refer to this variant.

\section{Experimental Evaluation}
\label{sec:experimental-evaluation}
Experimental evaluation consists of measuring the execution time, speedup, scaling, energy efficiency and memory usage of RTXRMQ and comparing these results with state of the art GPU/CPU approaches that made their source code available, as well as with a basic GPU approach for reference. In the RMQ tests, the input array is randomly generated as floats following a uniform distribution. Each query chooses the starting location of the $(l,r)$ range using a random uniform distribution, and the range length (\textit{i.e.}, $r-l+1$) is chosen using a random uniform or log-Normal distribution in order to represent large, medium or small range queries in relation to the input size $n$. More details on the range distributions are given in subsection \ref{subsec:performance-results}.
\subsection{Approaches}
The following approaches are considered for evaluation:
\begin{itemize}
    
    \item \textbf{RTXRMQ:} The proposed RT approach using the block-matrix extension from Subsection \ref{subsec:rtxrmq-blocks}. RTXRMQ uses Algorithms \ref{alg:gen_triang_blocks},\ref{alg:raygen_blocks}, and \ref{alg:closest_hit}. By being block-based it can reach inputs of size $n \ge 2^{24}$. The source code is public\footnote{RTXRMQ at \url{https://github.com/temporal-hpc/rtxrmq}.}.

    \item \textbf{HRMQ:} A state of the art CPU-based RMQ solution \cite{Ferrada2016ImprovedRM} with source code available\footnote{HRMQ at \url{https://github.com/hferrada/rmq}.}. Here, the implementation was slightly modified to solve several RMQs in parallel with OpenMP, without altering the core algorithm for each RMQ. With this change, HRMQ can scale its performance on multi-core CPUs. 
    
    \item \textbf{LCA:} A fast state of the art GPU-based RMQ solution, by Polak \textit{et al.} \cite{polak2021euler}, with source code available\footnote{LCA-GPU at \url{https://github.com/stobis/euler-meets-cuda}.}. It computes RMQs by answering the dual problem; Lowest Common Ancestor (LCA). LCA uses the \texttt{Modern GPU} library \cite{Baxter:2016:M2} which does computation with the regular GPU compute units (\textit{i.e.}, no use of RT Cores).

    \item \textbf{EXHAUSTIVE:} A basic CUDA GPU implementation, where each thread handles one query and searches the minimum from left to right in the $(l,r)$ range of the input array. This implementation serves as a reference on what is the parallel performance of a rapidly developed GPU solution with no special algorithmic improvements.  
\end{itemize}

\subsection{Hardware Setup}
The hardware setup used is detailed in  Table \ref{tab:hardware}.

\begin{table}[ht!]
\caption{GPU and system used for the experimentation.}
\label{tab:hardware}
\centering
\resizebox{0.9\columnwidth}{!}{
\begin{tabular}{|ll|l|}
\hline
\multicolumn{2}{|l|}{\textbf{\cellcolor{temporal}Attribute}} & \cellcolor{temporal}\textbf{RTX 6000 Ada}\\ \hline
\multicolumn{2}{|l|}{\textbf{Architecture}}                  & Lovelace (2022)                          \\ \hline
\multicolumn{2}{|l|}{\textbf{GPU Chip}}                      & AD102                                    \\ \hline
\multicolumn{2}{|l|}{\textbf{Memory}}                        & $48$ GB                                  \\ \hline
\multicolumn{2}{|l|}{\textbf{Cores (FP32)}}                  & $18176$                                  \\ \hline
\multicolumn{2}{|l|}{\textbf{SMs}}                           & $142$                                    \\ \hline
\multicolumn{2}{|l|}{\cellcolor{yellow}\textbf{RT Cores}}    & \cellcolor{yellow}$142$ (3rd Gen)        \\ \hline
\multicolumn{2}{|l|}{\textbf{Tensor Cores}}                  & $512$                                    \\ \hline
\multicolumn{2}{|l|}{\textbf{Cores/SM}}                      & $64$                                     \\ \hline
\multicolumn{2}{|l|}{\textbf{Mem. Bandwidth}}                & $960$ GB$/s$                             \\ \hline
\multicolumn{2}{|l|}{\textbf{Thermal Design Power (TDP)}}    & $300$W                                   \\ \hline
\multicolumn{1}{|l|}{\multirow{3}{*}[-0.5em]{\textbf{System}}} & \textbf{OS}  & Ubuntu 22.04 LTS        \\ \cline{2-3} 
\multicolumn{1}{|l|}{} & \textbf{CUDA version}                 & 12.1                                   \\\cline{2-3} 
\multicolumn{1}{|l|}{} & \textbf{CPU}                          & 2$\times$ 96-core AMD EPYC 9654        \\ \cline{2-3} 
\multicolumn{1}{|l|}{} & \textbf{RAM}                          & 384GB DDR5                             \\ \hline
\end{tabular}
}
\end{table}

\subsection{Profiling the Approaches}
A performance heat map, is generated for each approach in the $n \times |(l,r)|$ space, where $n$ is the input array size and $|(l,r)|$ is the query range length\footnote{We recall that the length $|(l,r)|$ of a query $\text{RMQ}(l,r)$ is defined as $r-l+1$.} as a fraction of $n$, that is $|(l,r)| = n2^y$ with $y$ a negative exponent. The purpose of the heat map is to visualize the fastest (towards blue) and slowest (towards yellow) regions of the approach given $n$ and $|(l,r)|$. For RTXRMQ, there is a 2D and a 3D heatmap. The 2D one corresponds to the projection to $n \times |(l,r)|$ space with the fastest block configuration, while the 3D heat map shows the entire configuration space where the optimal block configurations (regions towards blue) can be seen located at different parts of the cube. It is worth mentioning that the each heat map has its own color scale, thus color comparisons are only valid within the same approach.

\begin{figure*}[ht!]
\resizebox{\textwidth}{!}{
    \includegraphics[scale=1]{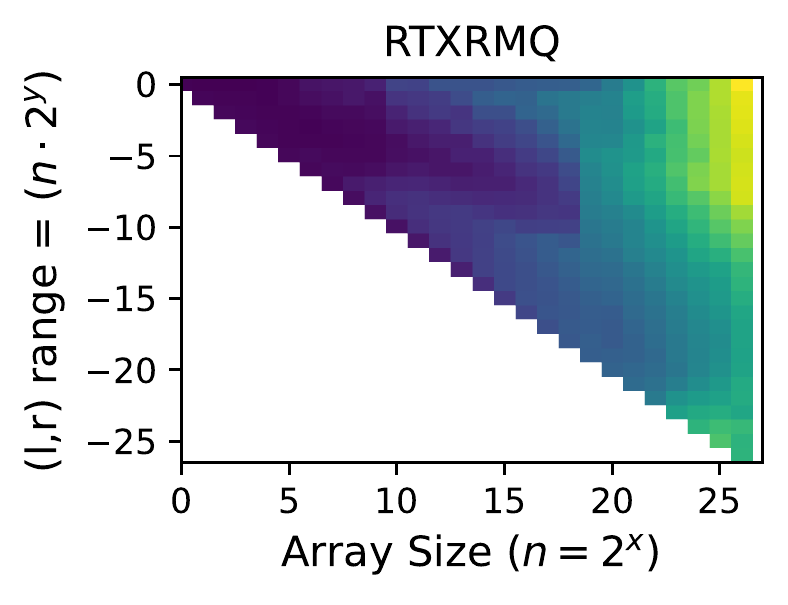}
    \includegraphics[scale=1]{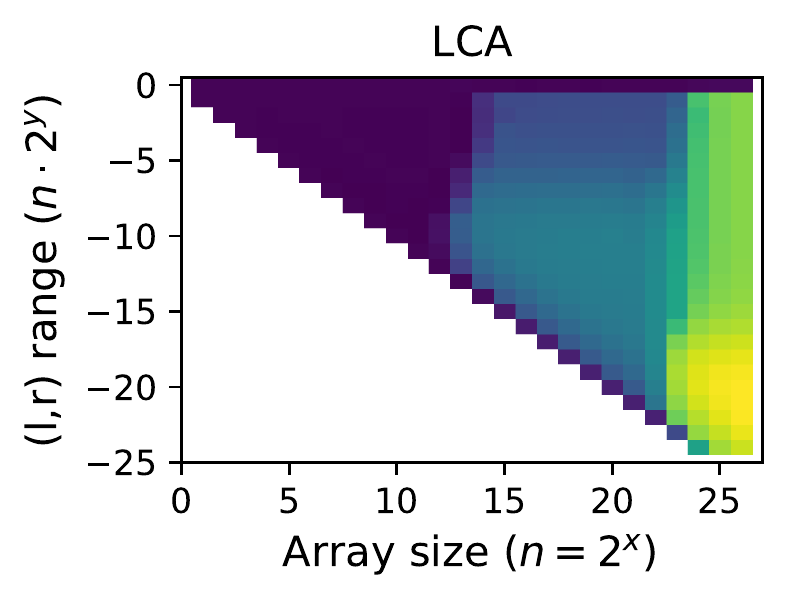}
    \includegraphics[scale=1]{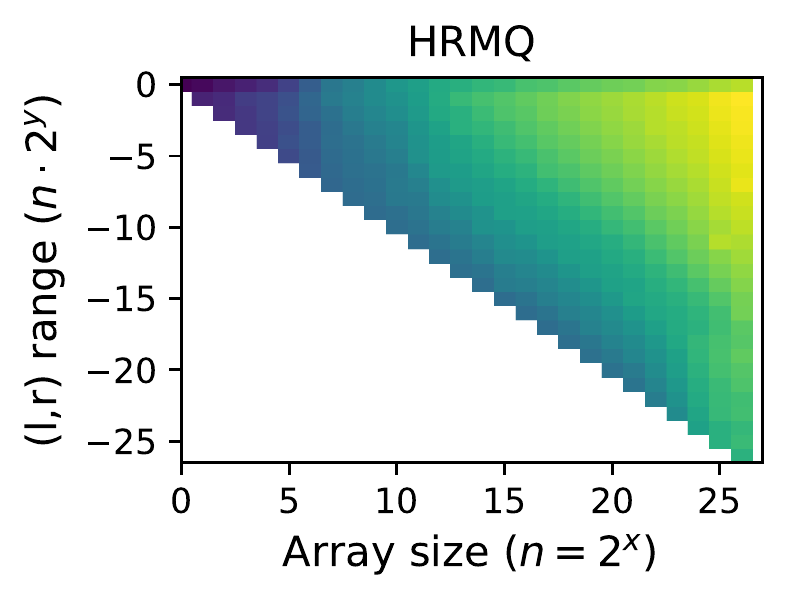}
    \includegraphics[scale=1]{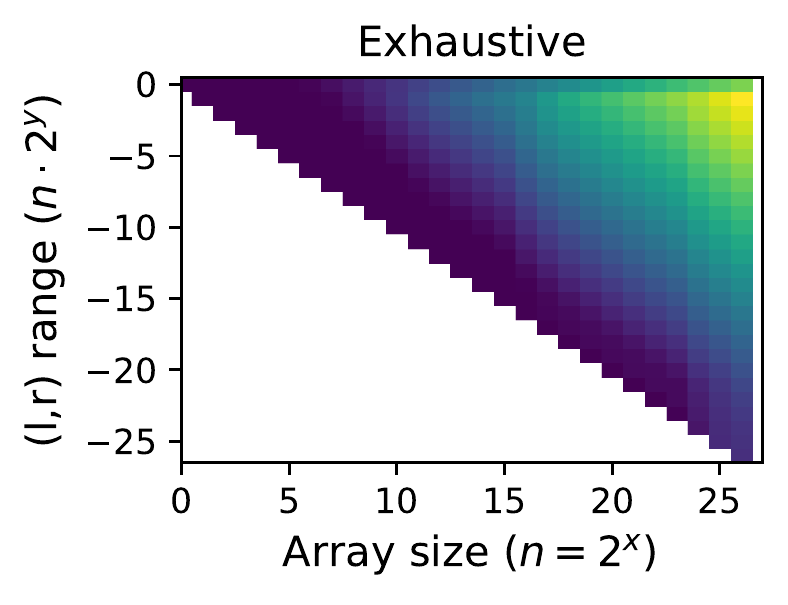}
}
\caption{Performance heat maps for each approach in $n \times |(l,r)|$ space, using the RTX 6000 Ada GPU and in the case of HRMQ, using 2$\times$ AMD EPYC 9654 CPUs (192 CPU cores). Blue (cold) regions correspond to faster performance while yellow (warm) to slower.}
\label{fig:profiling}
\end{figure*}

\begin{figure}[ht!]
\centering
\includegraphics[scale=0.18]{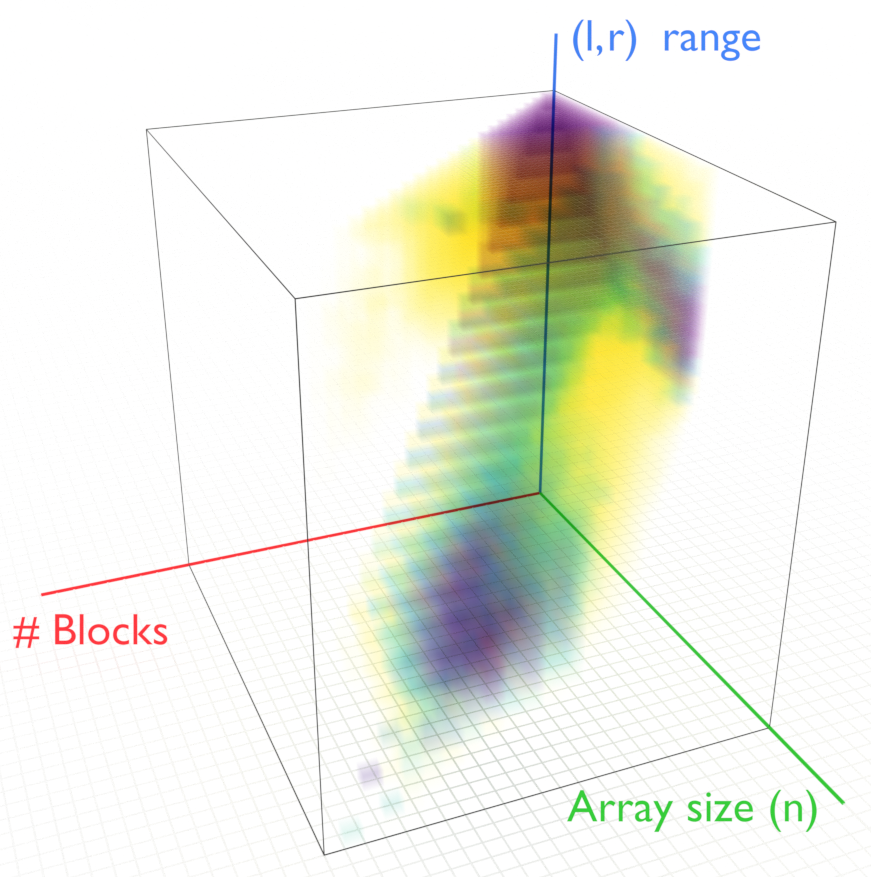}
\caption{3D heat map for RTXRMQ. Blue regions correspond to faster performance while yellow ones to slowest.}
\label{fig:hmap3D}
\end{figure}

Figure \ref{fig:profiling} shows the 2D heat maps of all approaches, where different performance behaviors can be observed. 
The first heat map is the 2D projection of RTXRMQ, where its cube (from Figure \ref{fig:hmap3D}) is projected to the $n,|(l,r)|$ plane by choosing optimal or near-optimal number of blocks at each $(n,|(l,r)|)$ location. This heat map exhibits two blue higher performance regions that extend horizontally across the $n$ axis. One of the regions is abruptly interrupted at the center of the heatmap; this is due to the numerical precision limitations of floating point in RT Cores, which required to filter the number of blocks to $nb < 2^{24}$ and the block size to $bs < 2^{18}$. Even with the filtered block configurations, the relevant behavior for this heat map is that for the large-$n$ regime, small/medium range queries run faster than large range ones. The second heat map presents LCA, with abrupt performance jumps at $n\sim 2^{13}$ and $n\sim 2^{22}$. In this case, at the large-$n$ regime the small/medium range queries run slower than long range ones. The third heat map presents HRMQ (CPU), where the gradient is smoother across the plane. Also for a same problem size $n$, small/medium range queries (bottom) run faster than large range ones (top). Lastly, the fourth heat map presents the Exhaustive approach, where small ranges always favor performance, for all $n$.

Figure \ref{fig:hmap3D} presents the 3D heat map of RTXRMQ with the non-valid configurations filtered out. This heat map shows that for a given $(n,|(l,r)|)$ tuple, performance can vary significantly depending on what number of blocks is chosen. Two high performance paths can be identified in the cube, one going down through the 3D diagonal that ends with a cluster of high performance, and a shorter path traveling through the $n,(l,r)$ plane that abruptly ends because of the filtered block configurations. 

\subsection{Time and Speedup}
\label{subsec:performance-results}
A performance benchmark was ran for each approach. The tests measure the time to solve a batch of $2^{26}$ RMQs on an array of $n$ randomly generated elements in normalized $[0,1]$ space. The metrics computed are the nanoseconds per RMQ ($ns/\text{RMQ}$) and the speedup over HRMQ. The queries are positioned randomly in the array with the $(l,r)$ range having one of the following distributions:
\begin{itemize}
    \item \textbf{Large $(l,r)$ range:} follows a random uniform distribution in $[1,n]$ with a mean of $(r-l) \approx n/2$.  
    \item \textbf{Medium $(l,r)$ range:} follows a Log-Normal distribution $\mathcal{LN}(\mu=\log(n^{0.6}), \sigma=0.3)$. As a reference, for $n=2^{26}$ the mean sits at $\sim 2^{15}$.
    \item \textbf{Small $(l,r)$ range:} follows a Log-Normal distribution $\mathcal{LN}(\mu=\log(n^{0.3}), \sigma=0.3)$. As a reference, at $n=2^{26}$ the mean sits at $\sim 2^8$. 
\end{itemize}
These three distributions are tested for each approach. For each value of $n$, the running time is computed as the average of 16 realizations\footnote{Except for the Exhaustive approach at large $n$, where two or three realizations  two internal repetitions sufficed due to the very high running times.}, where each realization is the average of 32 repeats. 

\begin{figure*}[ht!]
\resizebox{\textwidth}{!}{
    \includegraphics[scale=1.0]{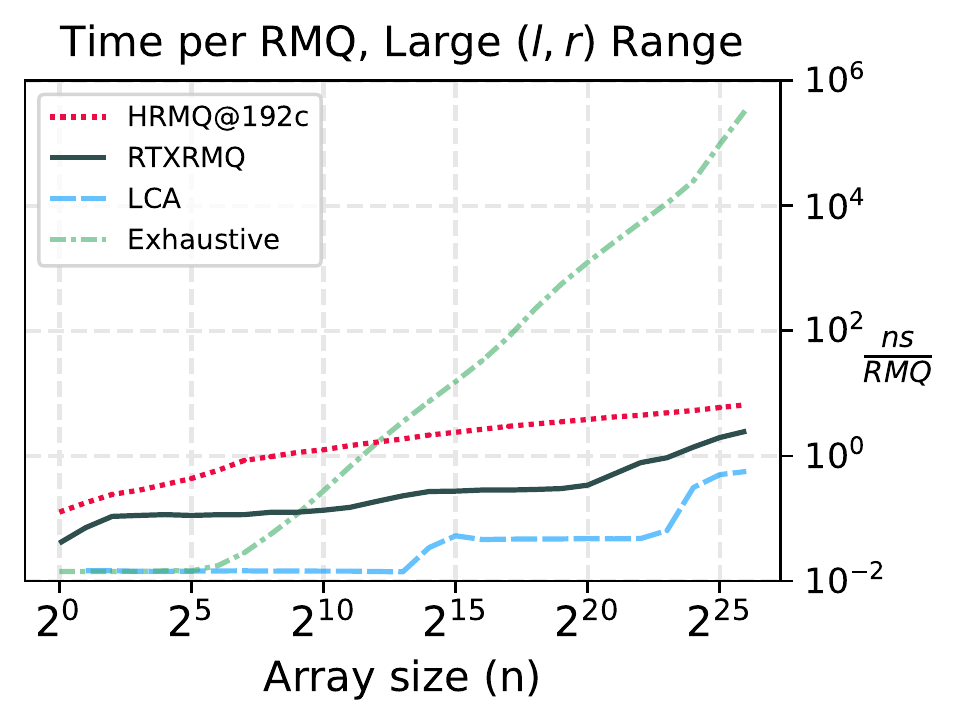}
    \includegraphics[scale=1.0]{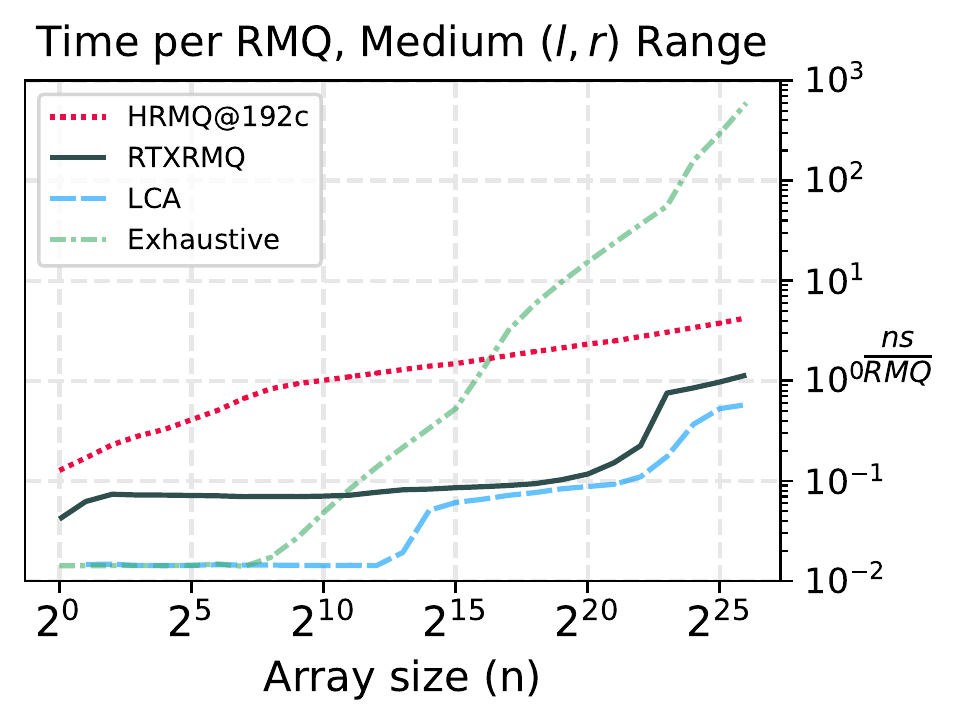}
    \includegraphics[scale=1.0]{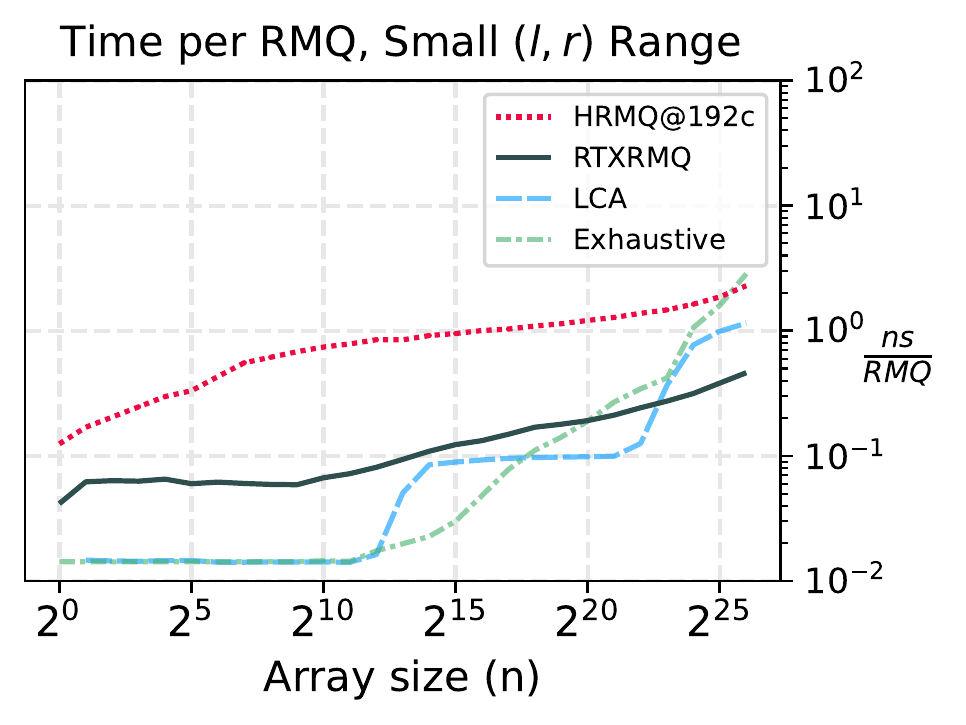}\\
}
\resizebox{\textwidth}{!}{
    \includegraphics[scale=1.0]{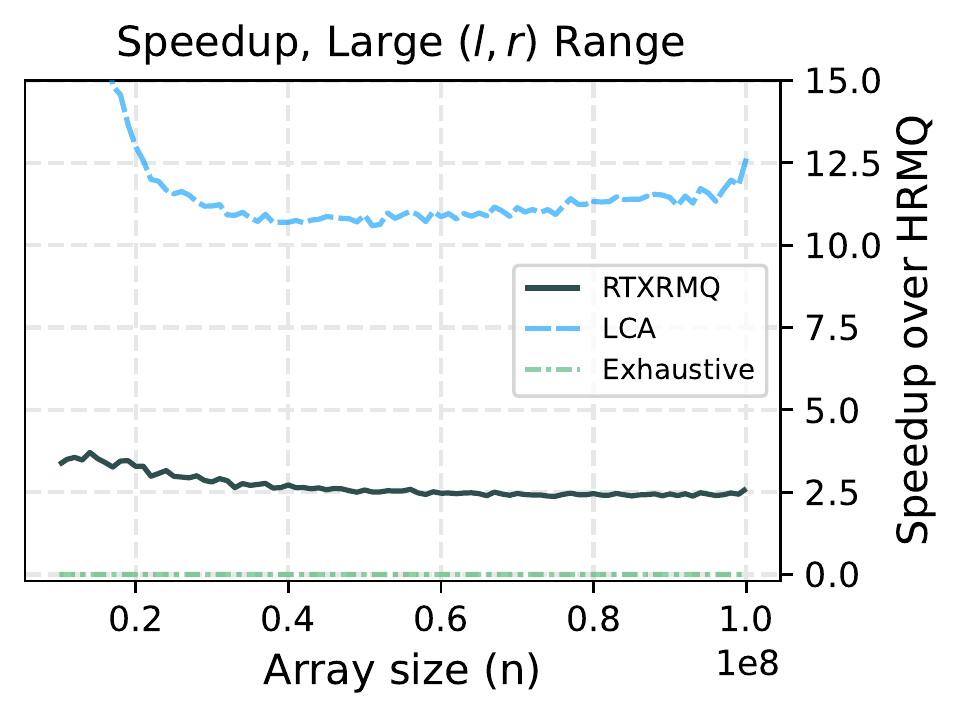}
    \includegraphics[scale=1.0]{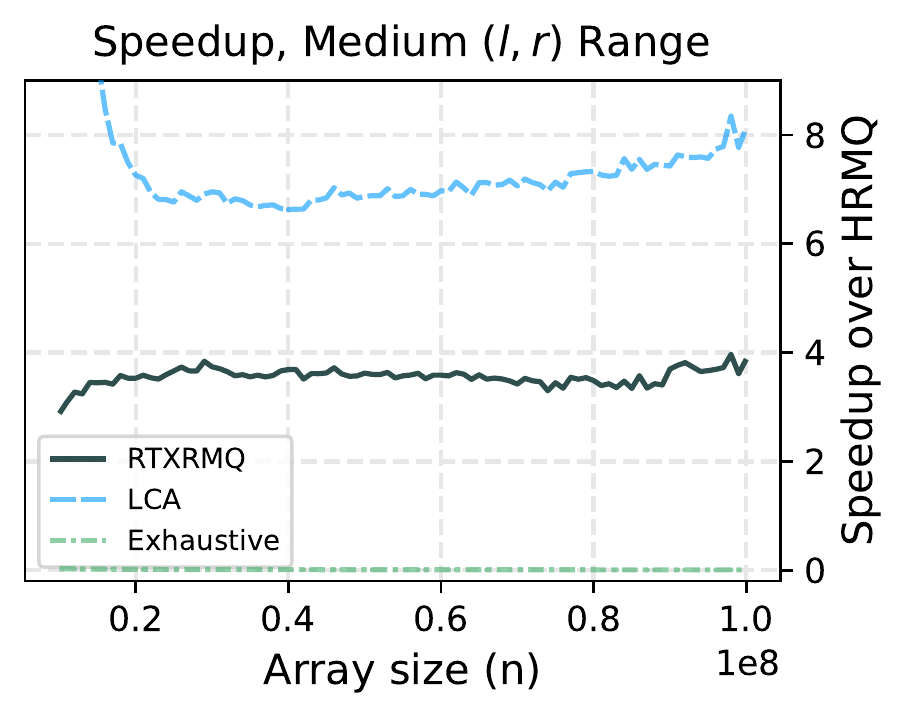}
    \includegraphics[scale=1.0]{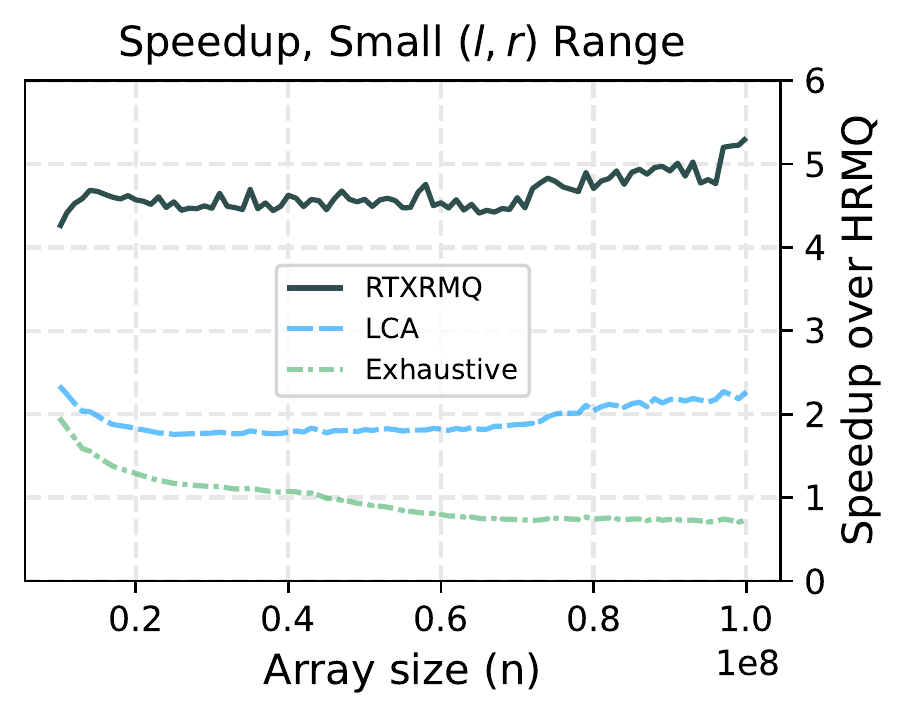}
}
\caption{Average time per RMQ and speedup over HRMQ under different $(l,r)$ range distributions.}
\label{fig:perf-time}
\end{figure*}

Figure \ref{fig:perf-time} presents the average time per query as nanoseconds per RMQ ($ns/RMQ$) for all approaches as well as the speedup of the GPU approaches over HRMQ which is a parallel CPU solution. The RTXRMQ, LCA and Exhaustive approaches run on the RTX 6000 Ada GPU and HRMQ runs on 2$\times$ AMD EPYC 9654 CPUs using a total of 192 CPU cores. The time per RMQ plots explore $n$ in powers of two, while the speedup plots focus on the largest problem sizes, exploring $n$ in the range of millions, \textit{i.e.}, $[10M\dots 100M]$. 
From the time per query plots, it can be seen that RTXRMQ runs competitive for large $n$, and in general exhibits a smooth increasing shape, except in the medium $(l,r)$ range at $n=2^{24}$. In the case of HRMQ, it also runs competitive (although slower than RTXRMQ and LCA) when being executed in parallel with 192 CPU cores. In the case of LCA, it exhibits a staircase shape in all three cases, which is consistent with the results of their authors, Polak \textit{et al.} \cite{polak2021euler}, where constant time behavior was claimed within a certain sub-range of $n$. Here, constant behavior behavior is not seen because these tests use a more extensive $n$ range, revealing that this constant time switches to different levels at certain problem sizes due to the effect of caches $L_1,L_2$ and VRAM in the GPU. It is worth mentioning that although the time per RMQ plots show that RTXRMQ is more competitive only in the range $n = 2^{22}\dots 2^{26}$, this range is exponentially larger than all the prior range $n < 2^{22}$. For this, the speedup plots explore the high $n$ range linearly.

In the large $(l,r)$ range (first column), RTXRMQ presents a performance of $\sim 5\ ns/RMQ$ for large $(l,r)$ ranges, which is $2.5\times$ faster than HRMQ running in parallel with 192 CPU cores. This speedup achieved by RTXRMQ, although significant, is greatly surpassed by LCA, which achieves up to $12.5\times$ of speedup, making RTXRMQ not the fastest approach for large $(l,r)$ ranges. It is worth mentioning that the Exhaustive approach is orders of magnitude slower than all other approaches, confirming that a brute force solution is still impractical even when ran in parallel with GPUs. 

In the medium $(l,r)$ range (second column), RTXRMQ still sits in between HRMQ and LCA, but narrows the performance difference with respect to LCA, as it now reaches $4\times$ of speedup, while LCA reaches $8\times$. Another aspect to highlight is that the $ns/RMQ$ improvement with respect to the large range is higher for HRMQ and RTXRMQ and less for LCA. For the case of the Exhaustive approach, once again it suffers from very slow performance, being orders of magnitude slower than the rest, but at the same time significantly faster than the large $(l,r)$ range.  

In the small $(l,r)$ range (third column), RTXRMQ becomes the most competitive approach once the problem size is large in the order of millions of elements. This is confirmed in the speedup plot, which shows that in the range from 10 to 100 millions of elements, RTXRMQ is the fastest approach reaching up to $5\times$ of speedup over HRMQ, and $2.3\times$ over LCA. The HRMQ approach also benefits from smaller queries, showing an improvement relative to the other cases. Lastly, the Exhaustive approach manages to be competitive in this tests, surprisingly being the fastest approach for a range of small problem sizes in the scale of $n \sim 2^{15}$. After this range, performance degrades to be the slowest in large problem sizes.

Although HRMQ has in general slower performance than RTXRMQ and LCA, it is highly competitive for being a parallel CPU solution, showing that it can scale its performance in a CPU server with 192 cores (two 96 core sockets) by parallelizing at the queries level. Also it shows an overall promising curve shape that can make it even more competitive for even larger input sizes not showed here because of memory constraints. In the case of RTXRMQ, it is the fastest for small ranges but surpassed by LCA in large and medium ones. Performance scaling analysis can provide further insights on knowing if RTXRMQ could match or surpass LCA in future GPUs.

\subsection{Scaling}
\begin{figure*}[ht!]
\resizebox{\textwidth}{!}{
    \includegraphics[scale=1.0]{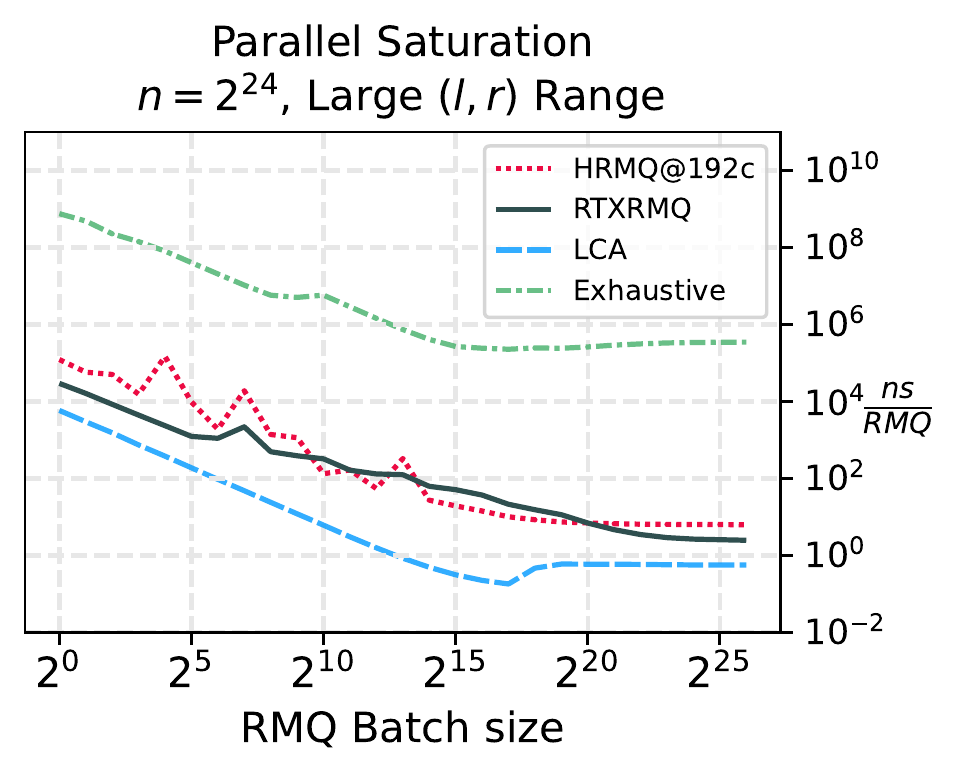}
    \includegraphics[scale=1.0]{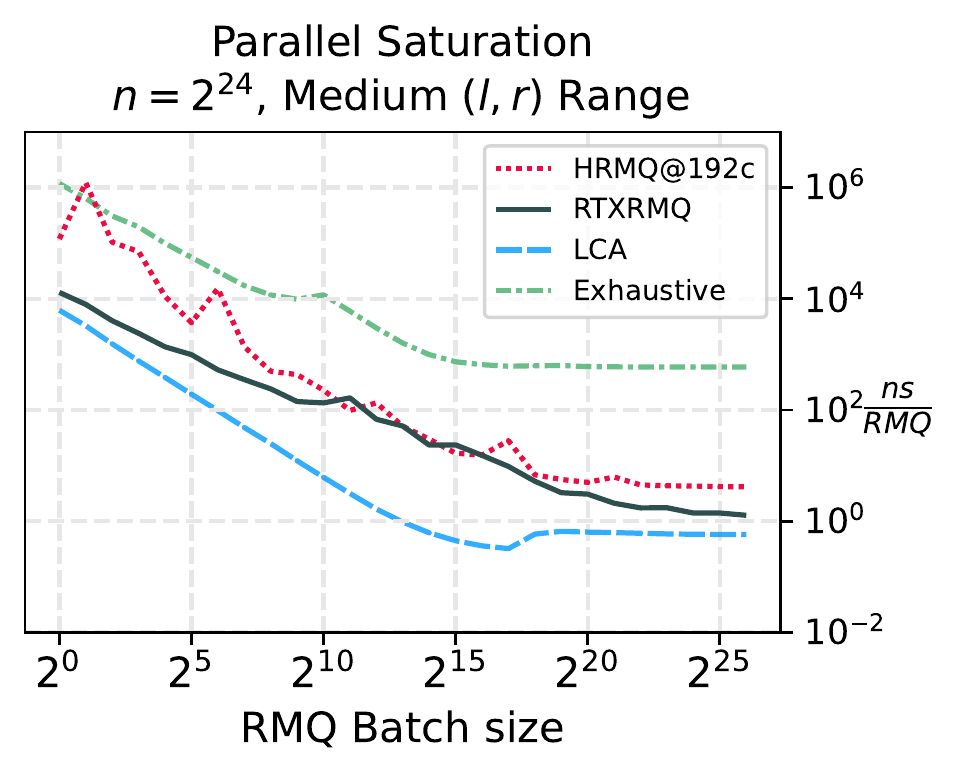}
    \includegraphics[scale=1.0]{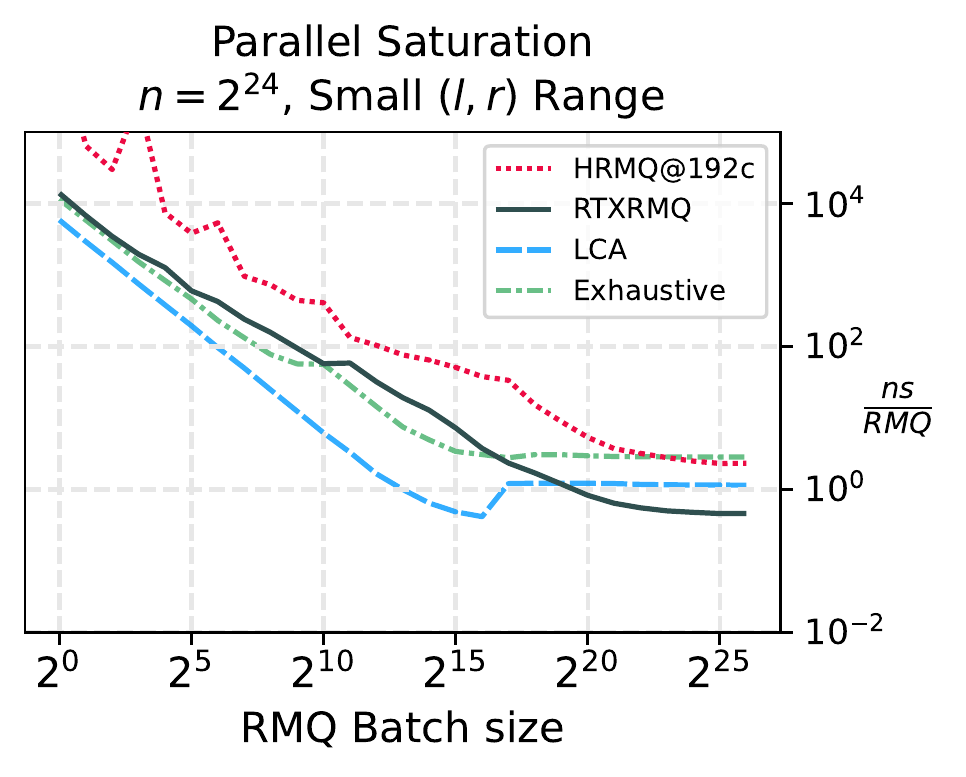}
}
\caption{Performance scaling as the RMQ Batch size increases.}
\label{fig:perf-vs-q}
\end{figure*}

The time and speedup plots from Figure \ref{fig:perf-time} used an RMQ batch size of $2^{26}$ queries as it is a relatively high amount of parallel work to test for each $n$ within a reasonable amount of time. The parallel saturation can provide additional insights on the scaling of each approach as the RMQ batch size increases. Figure \ref{fig:perf-vs-q} presents the parallel saturation, ranging the RMQ batch size from 1 to $2^{26}$, in order to know at which point parallel performance becomes saturated and stable.
From the plots one can note that all approaches, except RTXRMQ, saturate at some point near $2^{18}$ queries, with LCA having a particular degradation near $2^{17}$ that may be attributed to some of the data structures no longer fitting entirely in the L2 Cache (96 MB) of the GPU. In the case of RTXRMQ, it does not reach parallel saturation in the tested range. This can be interpreted as positive as increasing the batch size beyond $2^{26}$ would still provide additional performance to RTXRMQ, while the rest of the approaches would remain with the same performance. This opens the possibility for RTXRMQ to be the fastest approach once a sufficiently large RMQ batch size is used.

\begin{figure}[ht!]
    \centering
    \includegraphics[scale=0.65]{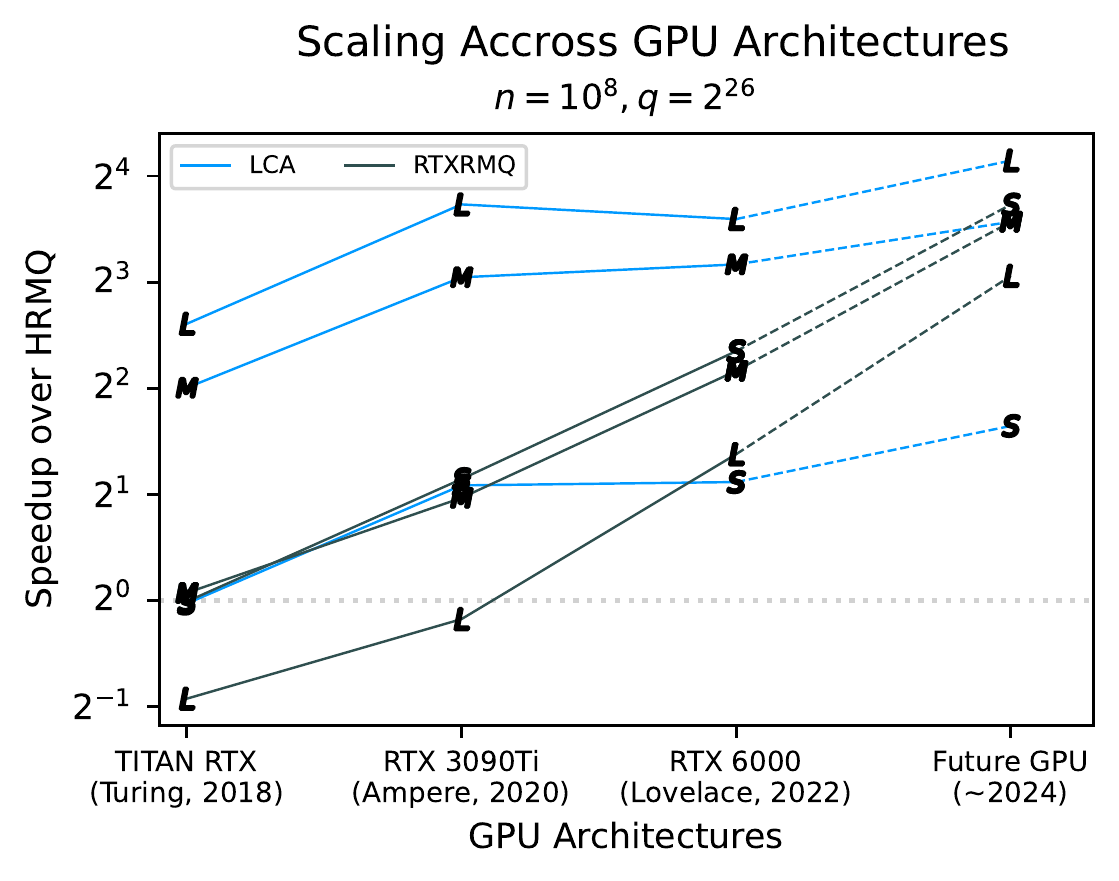}
    \caption{Performance scaling of RTXRMQ and LCA for Large (L), Medium (M) and Small (S) $(l,r)$ ranges, across different GPU architectures.}
    \label{fig:scaling-arch}
\end{figure}

Figure \ref{fig:scaling-arch} shows how RTXRMQ and LCA (the two most competitive approaches) scale their performance across consecutive GPU architectures. In this case, the architectures tested are Turing (2018, TITAN RTX GPU), Ampere (2020, RTX 3090Ti GPU) and Lovelace (2022, RTX 6000 Ada). Additionally, performance is projected for a future GPU architecture in the case that the observed trend for each approach continues.
From the plot, it is worth noticing that RTXRMQ is favored with near-exponential scaling ratios in these last three generations. On the other hand, LCA which runs with regular CUDA computation still scales performance, but not at the rate of RT Cores. The projection shows that in the case the trend continues, the next GPU architecture would narrow the performance difference with LCA in the large $(l,r)$ range, would make RTXRMQ faster than LCA in the medium $(l,r)$ range, and would further increase the speedup of RTXRMQ over LCA in the small $(l,r)$ range.

\begin{figure}[ht!]
    \centering
    \includegraphics[scale=0.65]{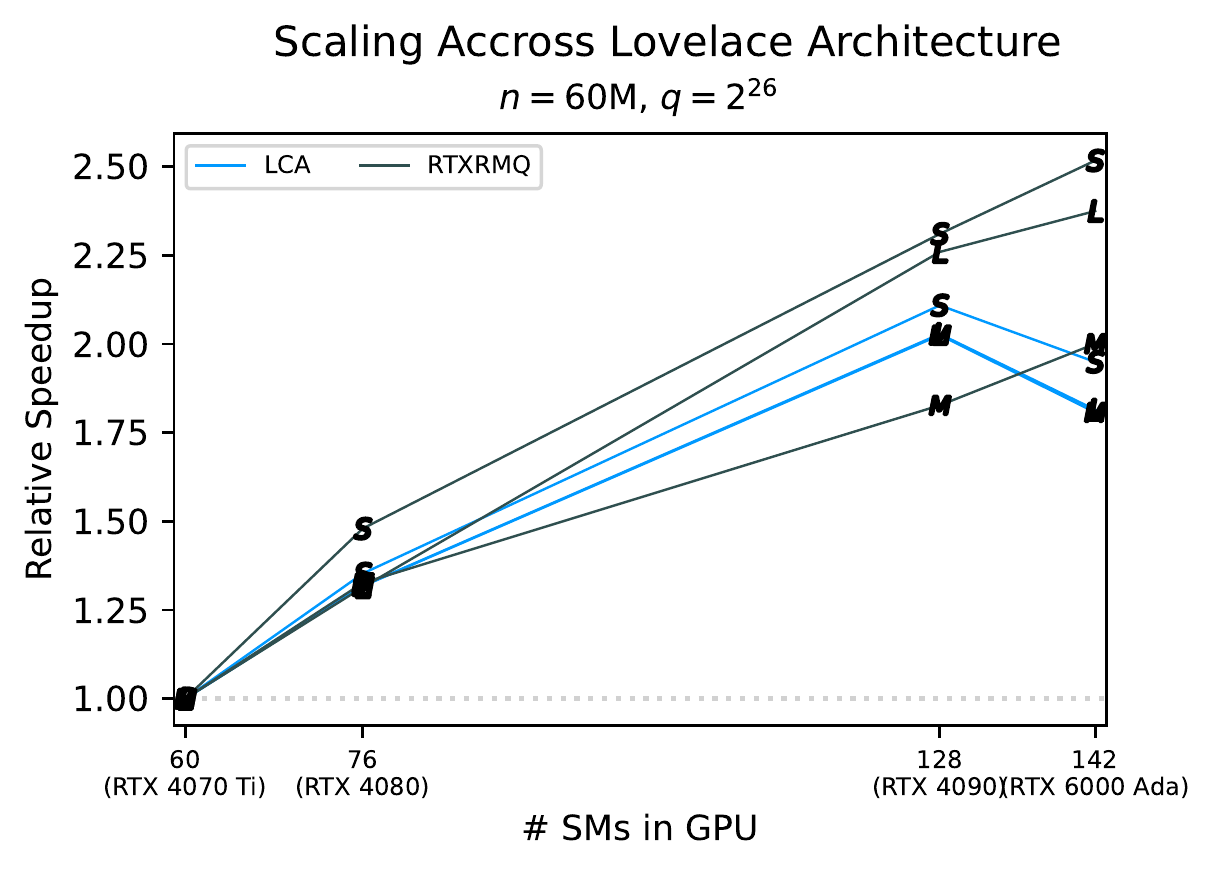}
    \caption{Scaling of RTXRMQ and LCA for Large (L), Medium (M) and Small (S) $(l,r)$ ranges.}
    \label{fig:scaling-sm}
\end{figure}

Figure \ref{fig:scaling-sm} presents the scaling of RTXRMQ and LCA within the same GPU architecture (Lovelace, 2022), but varying the number of streaming multiprocessors. This was accomplished by testing the approaches on four different GPUs of the Lovelace generation; RTX 4070 Ti (60 SMs), RTX 4080 (76 SMs), RTX 4090 (128 SMs) and RTX 6000 Ada (142 SMs).

From the plot, RTXRMQ manages to scale RT Core performance practically linear to the number of SMs, while LCA only scales linear up to the RTX 4090, but then scales down when using the 142 SMs of the RTX 6000 Ada GPU.

\subsection{Energy Efficiency}
\begin{figure*}[ht!]
\resizebox{\textwidth}{!}{
    \includegraphics[scale=1.0]{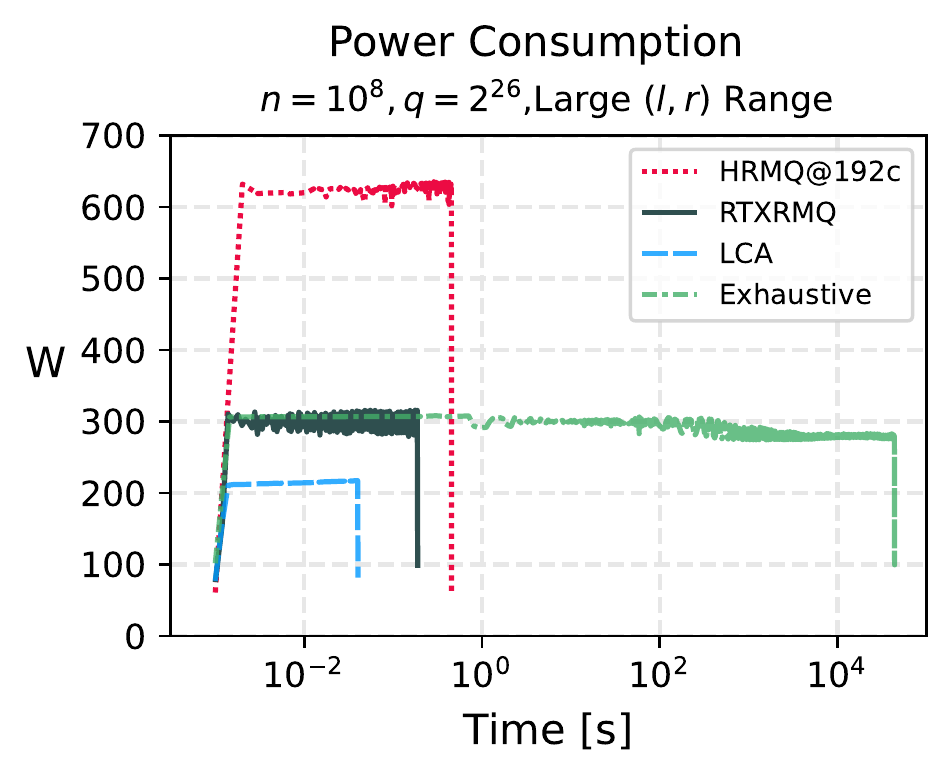}
    \includegraphics[scale=1.0]{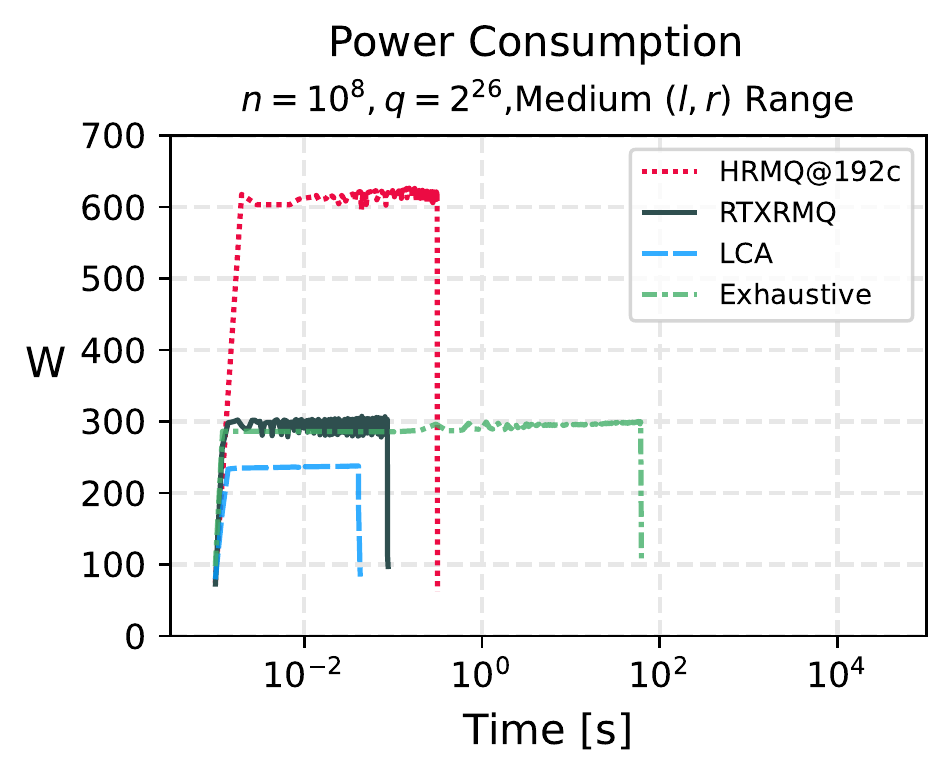}
    \includegraphics[scale=1.0]{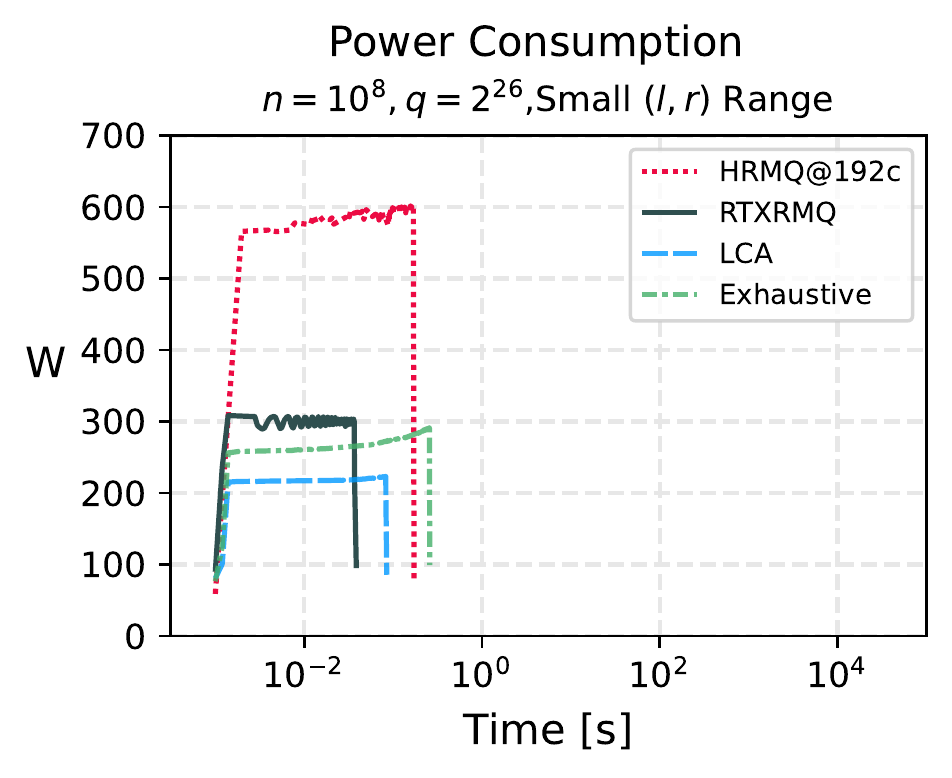}
}
\caption{Power time series for all approaches, under the three different distributions; Large, Medium and Small $(l,r)$ ranges.}
\label{fig:power}
\end{figure*}

Power consumption and number of RMQs per Joule were measured for all approaches running a test of $n=10^8$ elements, batch of $q=2^{26}$ RMQs, under large, medium and small $(l,r)$ ranges. Figure \ref{fig:power} presents the power time series of all approaches.

\begin{figure}[ht!]
    \centering
    \includegraphics[scale=0.65]{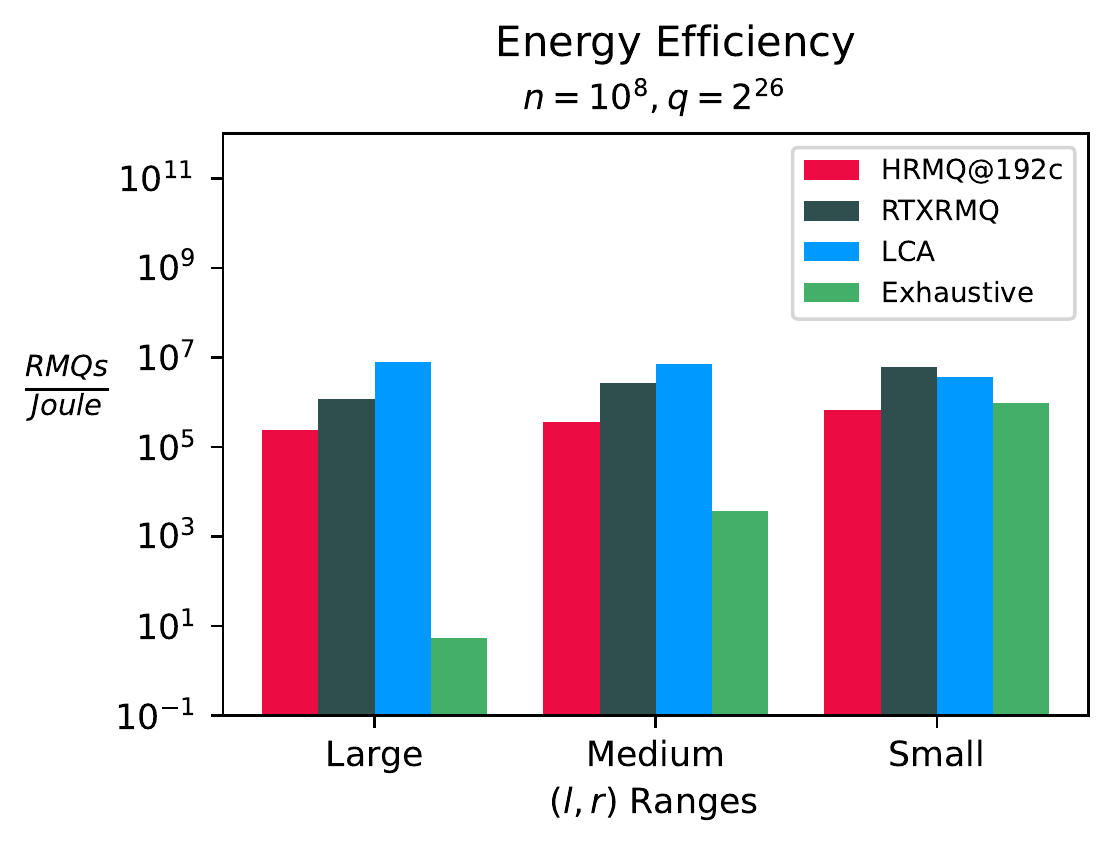}
    \caption{Energy efficiency under large, medium and small $(l,r)$ ranges.}
    \label{fig:energy-efficiency}
\end{figure}

The time series show that all approaches exhibit a stable power consumption along their execution. In terms of peak power consumption, for large and medium $(l,r)$ ranges (first and second plots), RTXRMQ and Exhaustive reach 300W which matches the thermal design power (TDP) of the RTX 6000 Ada GPU, while LCA consumes between 200W and 240W along its execution. On the other hand, the CPU approach, HRMQ, consumes up to 600W during execution which is close to the resulting TDP of 720W from the two AMD EPYC 9654 CPUs used, with 96 cores each. In general, the less power consuming approach is LCA, followed by RTXRMQ and Exhaustive. In the small $(l,r)$ range, even when RTXRMQ consumes more Watts than LCA, it compensates by taking less time, leading to a superior energy efficiency. Figure \ref{fig:energy-efficiency} presents the average number of RMQs per Joule of energy for all approaches using the same input sizes $n=10^8M, q=2^{26}$, under the three different $(l,r)$ range distributions.

The plot shows that in large and medium $(l,r)$ ranges, LCA is the most energy efficient approach, followed by RTXRMQ, and for small $(l,r)$ ranges, RTRMQ is the most energy efficient one. The HRMQ approach, although consumed 600W during execution, is still energy efficient following LCA and RTXRMQ in all $(l,r)$ ranges. Lastly, the Exhaustive approach shows in a clear way how it is not energy efficient for large and medium ranges, but rapidly improves its energy efficiency by orders of magnitude as the $(l,r)$ range decreases from large to small.

\subsection{Memory usage by the Data Structure}
Table \ref{tab:mem-usage} presents the memory usage (in MBytes) of each approach from the data structures required to operate.

\begin{table}[ht!]
\centering
\caption{Sizes in MB of the data structures of each approach. The compressed RTXRMQ case includes the percent of the default uncompressed BVH. The Exhaustive approach is excluded as it does not require a data structure.}
\label{tab:mem-usage}
\resizebox{\columnwidth}{!}{
\begin{tabular}{|r|rr|r|r|}
\hline
\rowcolor[HTML]{70BAA4} 
\multicolumn{1}{|c|}{\cellcolor[HTML]{70BAA4}}                                      & \multicolumn{2}{c|}{\cellcolor[HTML]{70BAA4}\textbf{RTXRMQ}}                                                                     & \multicolumn{1}{c|}{\cellcolor[HTML]{70BAA4}}                               & \multicolumn{1}{c|}{\cellcolor[HTML]{70BAA4}}                                \\ \cline{2-3}
\rowcolor[HTML]{70BAA4} 
\multicolumn{1}{|c|}{\multirow{-2}{*}{\cellcolor[HTML]{70BAA4}\textbf{Input Size}}} & \multicolumn{1}{c|}{\cellcolor[HTML]{70BAA4}\textbf{Default}} & \multicolumn{1}{c|}{\cellcolor[HTML]{70BAA4}\textbf{Compressed}} & \multicolumn{1}{c|}{\multirow{-2}{*}{\cellcolor[HTML]{70BAA4}\textbf{LCA}}} & \multicolumn{1}{c|}{\multirow{-2}{*}{\cellcolor[HTML]{70BAA4}\textbf{HRMQ}}} \\ \hline
\begin{tabular}[c]{@{}r@{}}$n=2^{10}$ \\ (0.004 MB)\end{tabular}                    & \multicolumn{1}{r|}{0.07 MB}                                  & 0.06 MB (85\%)                                                   & 0.334 MB                                                                    & 0.003 MB                                                                     \\ \hline
\begin{tabular}[c]{@{}r@{}}$n=2^{15}$\\ (0.131 MB)\end{tabular}                     & \multicolumn{1}{r|}{2.24 MB}                                  & 1.77 MB (79\%)                                                   & 0.55 MB                                                                     & 0.01 MB                                                                      \\ \hline
\begin{tabular}[c]{@{}r@{}}$n=2^{20}$\\ (4.19 MB)\end{tabular}                      & \multicolumn{1}{r|}{71.63 MB}                                 & 56.28 MB (78\%)                                                  & 6.93 MB                                                                     & 0.30 MB                                                                      \\ \hline
\begin{tabular}[c]{@{}r@{}}$n=2^{26}$\\ (268.43 MB)\end{tabular}                    & \multicolumn{1}{r|}{4512.15 MB}                               & 3601.46 MB (79\%)                                                & 170.52 MB                                                                   & 20.12 MB                                                                     \\ \hline
\end{tabular}
}
\end{table}

From the Table, HRMQ is clearly the less memory consuming approach. In the largest case, $n=2^{26}$, HRMQ requires only 20MB of memory to handle its data structures which is orders of magnitude smaller than RTXRMQ and LCA. This is due to the efficient compact data structure employed internally. Following HRMQ comes LCA using an order of magnitude more of extra memory, reaching up to 170 MB for the largest case. Lastly, RTXRMQ is the most memory consuming approach, with up to 4.5 GB of memory using the BVH in its default uncompacted form, and up to 3.6 GB when using OptiX's BVH compaction\footnote{The performance of RTXRMQ was not affected by the compaction option.} feature. This high consumption is necessary in order to build a geometrical model for RMQ, because the BVH needs to store triangles composed of three vertices, with each vertex having three coordinates. This translates to the BVH having to handle at least $9n$ the original size, with $n$ the number of elements of the input array. Nevertheless, it is worth mentioning that RTXRMQ can answer an RMQ by both index or value, whereas HRMQ and LCA can only answer through the index. 

\section{Discussion and Conclusions}
\label{sec:conclusions}
The present work has presented RTXRMQ, an approach for solving batches of range minimum queries (RMQs) with GPU Ray Tracing (RT) Cores. The key idea is to transform the input array to a set of $YZ$ aligned triangles along the $X$-axis according to their value, with their shapes along $Y$ and $Z$ according to their index in the array. With this new model, any $\text{RMQ}(l,r)$ is now solved by launching a ray from the $YZ$ plane along the $X$-axis, and the closest hit with a triangle corresponds to the minimum element in the range $(l,r)$, answering $\text{RMQ}(l,r)$. To overcome 32-bit floating point precision limitations of OptiX, RTXRMQ uses a block matrix approach, where instead of handling all queries in one normalized space, the input array is distributed into blocks of sub-spaces arranged in the $YZ$ plane. 

Experimental evaluation has shown that RTXRMQ is highly competitive when compared to state of the art approaches such as HRMQ  \cite{Ferrada2016ImprovedRM} and LCA \cite{polak2021euler}, which run on CPU (multi-core) and GPU respectively. More specifically, results have shown that RTXRMQ achieves a speedup over HRMQ of $2.5\times$, $4\times$ and $5\times$ for large, medium and small $(l,r)$ range distributions, respectively. When compared to LCA, the results are mixed; RTXRMQ is up to $2.3\times$ faster than LCA in the small $(l,r)$ range distribution but slower for large and medium ones. The reason why RTXRMQ does not perform as fast in these cases may be attributed to the fact that rays launched for large $(l,r)$ ranges require more internal BVH traversal because more internal nodes of the acceleration structure are being intersected. Nevertheless, RTXRMQ exhibits the strongest performance scaling ratio in both architectural jump as well as within an architecture. If such trend continues for the upcoming GPU architecture, then the projected performance would make RTXRMQ become faster than LCA for medium $(l,r)$ ranges, leaving LCA only faster in the large $(l,r)$ range and by a smaller margin. In addition, the parallel saturation plots showed evidence that RTXRMQ can further scale its performance on even larger RMQ batch sizes, whereas HRMQ and LCA would not as they already reached their parallel saturation point.
In terms of energy efficiency, RTXRMQ can be considered an energy efficient approach, only being surpassed by LCA in large and medium $(l,r)$ ranges, but significantly more energy efficient than HRMQ and an exhaustive GPU implementation used for base reference. In terms of memory usage, RTXRMQ uses a high amount of memory compared to the other approaches, something that may limit problem sizes to the current memory of the GPU. On the other hand, this higher memory consumption compensates by making RTXRMQ the only approach capable of answering by index (\textit{i.e.}, element position in the array) and by value, whereas HRMQ and LCA can only answer by index.

Future work can explore three possible aspects; \textbf{i)} to handle multiple BVHs such that each one handles a cluster of blocks, and evaluate new spatial organizations of them. This modification could provide performance benefits to RTXRMQ, as entire BVHs could be discarded when launching rays. A preliminary version of this optimization was attempted using one BVH per block, however performance was not superior than the current implementation with one BVH. It is possible that using one BVH for all blocks, and using one BVH per block, are both extreme design decisions, thus choosing an intermediate number of BVHs with a proper spatial distribution remains open. The other possible optimization is \textbf{ii)} to employ the new shader execution reordering (SER) feature recently included, but still not available in OptiX as of May 2023. Although the main feature of SER is the possibility to reorganize threads to achieve a less divergent execution of warps, it also offers the possibility to capture intersections with objects without needing to execute a closest hit or miss shader. This last aspect could allow RTXRMQ to detect intersections in a more anticipated way, with a possible improvement in performance. Lastly, \textbf{iii)} to explore dynamic RMQs, that is, being able to solve batches of RMQs for input arrays that change their values over time; useful for scientific applications such as simulations. RT cores offer fast update/rebuild functions that when used in a balanced way, may lead to an efficient solution. Research on this feature would open new possibilities for applying RMQs, and would justify the higher memory consumption in the BVH as it could handle dynamic scenarios that static approaches could not.

The results obtained in this work have shown that adapting a non-graphical problem to the RT core model, such as the range minimum query (RMQ), can provide significant performance benefits, efficient performance scaling and give new insights to the understanding of how RT cores should be employed to further accelerate new tasks in engineering and technology, including dynamic ones.

\section*{Acknowledgement}
This work was supported by the ANID FONDECYT grant \#1221357, the Temporal research group and the Patagón supercomputer of Universidad Austral de Chile (FONDEQUIP EQM180042). Special thanks to AMAX Engineering (\url{https://www.amax.com/}) for supporting with the CPU and GPU hardware used for the experimentation.

\bibliographystyle{elsarticle-num}
\bibliography{main}

\begin{thebibliography}{10}
\expandafter\ifx\csname url\endcsname\relax
  \def\url#1{\texttt{#1}}\fi
\expandafter\ifx\csname urlprefix\endcsname\relax\def\urlprefix{URL }\fi
\expandafter\ifx\csname href\endcsname\relax
  \def\href#1#2{#2} \def\path#1{#1}\fi

\bibitem{navarro2014survey}
C.~A. Navarro, N.~Hitschfeld-Kahler, L.~Mateu, A survey on parallel computing
  and its applications in data-parallel problems using {GPU} architectures,
  Communications in Computational Physics 15~(2) (2014) 285--329.

\bibitem{nickolls2010gpu}
J.~Nickolls, W.~J. Dally, The {GPU} computing era, IEEE micro 30~(2) (2010)
  56--69.

\bibitem{owens2008gpu}
J.~D. Owens, M.~Houston, D.~Luebke, S.~Green, J.~E. Stone, J.~C. Phillips,
  {GPU} computing, Proceedings of the IEEE 96~(5) (2008) 879--899.

\bibitem{nickolls2008scalable}
J.~Nickolls, I.~Buck, M.~Garland, K.~Skadron, Scalable parallel programming
  with cuda, Queue 6~(2) (2008) 40--53.

\bibitem{carrasco2018analyzing}
R.~Carrasco, R.~Vega, C.~A. Navarro, Analyzing {GPU} tensor core potential for
  fast reductions, in: 2018 37th International Conference of the Chilean
  Computer Science Society (SCCC), IEEE, 2018, pp. 1--6.

\bibitem{navarro2020gpu}
C.~A. Navarro, R.~Carrasco, R.~J. Barrientos, J.~A. Riquelme, R.~Vega, {GPU}
  tensor cores for fast arithmetic reductions, IEEE Transactions on Parallel
  and Distributed Systems 32~(1) (2020) 72--84.

\bibitem{dakkak2019accelerating}
A.~Dakkak, C.~Li, J.~Xiong, I.~Gelado, W.-m. Hwu, Accelerating reduction and
  scan using tensor core units, in: Proceedings of the ACM International
  Conference on Supercomputing, 2019, pp. 46--57.

\bibitem{NAVARRO2020158}
C.~A. Navarro, F.~A. Quezada, N.~Hitschfeld, R.~Vega, B.~Bustos, Efficient
  {GPU} thread mapping on embedded 2d fractals, Future Generation Computer
  Systems 113 (2020) 158 -- 169.

\bibitem{QUEZADA202210}
F.~A. Quezada, C.~A. Navarro, N.~Hitschfeld, B.~Bustos, Squeeze: Efficient
  compact fractals for tensor core {GPU}s, Future Generation Computer Systems
  135 (2022) 10--19.

\bibitem{sorna2018optimizing}
A.~Sorna, X.~Cheng, E.~D'azevedo, K.~Won, S.~Tomov, Optimizing the fast fourier
  transform using mixed precision on tensor core hardware, in: 2018 IEEE 25th
  International Conference on High Performance Computing Workshops (HiPCW),
  IEEE, 2018, pp. 3--7.

\bibitem{durrani2021fft}
S.~Durrani, M.~S. Chughtai, A.~Dakkak, W.-m. Hwu, L.~Rauchwerger, Fft blitz:
  the tensor cores strike back, in: Proceedings of the 26th ACM SIGPLAN
  Symposium on Principles and Practice of Parallel Programming, 2021, pp.
  488--489.

\bibitem{li2021tcfft}
B.~Li, S.~Cheng, J.~Lin, tcfft: A fast half-precision fft library for nvidia
  tensor cores, in: 2021 IEEE International Conference on Cluster Computing
  (CLUSTER), IEEE, 2021, pp. 1--11.

\bibitem{liu2022toward}
X.~Liu, Y.~Liu, H.~Yang, J.~Liao, M.~Li, Z.~Luan, D.~Qian, Toward accelerated
  stencil computation by adapting tensor core unit on {GPU}, in: Proceedings of
  the 36th ACM International Conference on Supercomputing, 2022, pp. 1--12.

\bibitem{zhu2022rtnn}
Y.~Zhu, {RTNN}: accelerating neighbor search using hardware ray tracing, in:
  Proceedings of the 27th ACM SIGPLAN Symposium on Principles and Practice of
  Parallel Programming, 2022, pp. 76--89.

\bibitem{zhao2023leveraging}
S.~Zhao, Z.~Lai, J.~Zhao, Leveraging ray tracing cores for particle-based
  simulations on {GPU}s, International Journal for Numerical Methods in
  Engineering 124~(3) (2023) 696--713.

\bibitem{zellmann2020accelerating}
S.~Zellmann, M.~Weier, I.~Wald, Accelerating force-directed graph drawing with
  {RT} cores, in: 2020 IEEE Visualization Conference (VIS), 2020, pp. 96--100.

\bibitem{morrical2020accelerating}
N.~Morrical, I.~Wald, W.~Usher, V.~Pascucci, Accelerating unstructured mesh
  point location with {RT} cores, IEEE Transactions on Visualization and
  Computer Graphics 28~(8) (2020) 2852--2866.

\bibitem{wald2019rtx}
I.~Wald, W.~Usher, N.~Morrical, L.~Lediaev, V.~Pascucci, {RTX} beyond ray
  tracing: Exploring the use of hardware ray tracing cores for tet-mesh point
  location., in: High Performance Graphics (Short Papers), 2019, pp. 7--13.

\bibitem{fischer2011space}
J.~Fischer, V.~Heun, Space-efficient preprocessing schemes for range minimum
  queries on static arrays, SIAM Journal on Computing 40~(2) (2011) 465--492.

\bibitem{fischer2006theoretical}
J.~Fischer, V.~Heun, Theoretical and practical improvements on the
  {RMQ}-problem, with applications to lca and lce, in: Annual Symposium on
  Combinatorial Pattern Matching, Springer, 2006, pp. 36--48.

\bibitem{muthukrishnan2002efficient}
S.~Muthukrishnan, Efficient algorithms for document retrieval problems., in:
  SODA, Vol.~2, Citeseer, 2002, pp. 657--666.

\bibitem{croft2010search}
W.~B. Croft, D.~Metzler, T.~Strohman, Search engines: Information retrieval in
  practice, Vol. 520, Addison-Wesley Reading, 2010.

\bibitem{kobayashi2000information}
M.~Kobayashi, K.~Takeda, Information retrieval on the web, ACM Computing
  Surveys (CSUR) 32~(2) (2000) 144--173.

\bibitem{abouelhoda2004chainer}
M.~I. Abouelhoda, E.~Ohlebusch, Chainer: Software for comparing genomes, in:
  Proceedings of the 12th International Conference on Intelligent Systems for
  Molecular Biology+ 3rd European Conference on Computational Biology,
  Citeseer, 2004.

\bibitem{bender2000lca}
M.~A. Bender, M.~Farach-Colton, The lca problem revisited, in: Latin American
  Symposium on Theoretical Informatics, Springer, 2000, pp. 88--94.

\bibitem{parker2010optix}
S.~G. Parker, J.~Bigler, A.~Dietrich, H.~Friedrich, J.~Hoberock, D.~Luebke,
  D.~McAllister, M.~McGuire, K.~Morley, A.~Robison, et~al., Optix: a general
  purpose ray tracing engine, Acm transactions on graphics (tog) 29~(4) (2010)
  1--13.

\bibitem{Ferrada2016ImprovedRM}
H.~Ferrada, G.~Navarro, Improved range minimum queries, 2016 Data Compression
  Conference (DCC) (2016) 516--525.

\bibitem{polak2021euler}
A.~Polak, A.~Siwiec, M.~Stobierski, Euler meets {GPU}: Practical graph
  algorithms with theoretical guarantees, in: 2021 IEEE International Parallel
  and Distributed Processing Symposium (IPDPS), IEEE, 2021, pp. 233--244.

\bibitem{Navarro2016}
G.~Navarro, Compact Data Structures: A Practical Approach, Cambridge University
  Press, 2016.

\bibitem{NM07}
G.~Navarro, V.~M\"{a}kinen, Compressed full-text indexes, ACM Comput. Surv.
  39~(1) (2007) 2–es.

\bibitem{BDMRRR05}
D.~Benoit, E.~D. Demaine, J.~I. Munro, R.~Raman, V.~Raman, S.~S. Rao,
  Representing trees of higher degree, Algorithmica 43~(4) (2005) 275–292.

\bibitem{Vu80}
J.~Vuillemin, A unifying look at data structures, Commun. ACM 23~(4) (1980)
  229–239.

\bibitem{FN17}
H.~Ferrada, G.~Navarro, Improved range minimum queries, Journal of Discrete
  Algorithms 43 (2017) 72--80.

\bibitem{AHU73}
A.~V. Aho, J.~E. Hopcroft, J.~D. Ullman, On finding lowest common ancestors in
  trees, in: Proceedings of the Fifth Annual ACM Symposium on Theory of
  Computing, STOC '73, Association for Computing Machinery, New York, NY, USA,
  1973, p. 253–265.

\bibitem{wald2007fast}
I.~Wald, On fast construction of sah-based bounding volume hierarchies, in:
  2007 IEEE Symposium on Interactive Ray Tracing, IEEE, 2007, pp. 33--40.

\bibitem{klosowski1998efficient}
J.~T. Klosowski, M.~Held, J.~S. Mitchell, H.~Sowizral, K.~Zikan, Efficient
  collision detection using bounding volume hierarchies of k-dops, IEEE
  transactions on Visualization and Computer Graphics 4~(1) (1998) 21--36.

\bibitem{NvidiaDeveloperBlog}
T.~Karras, Thinking in parallel, part ii, tree traversal on {GPU}, {Accessed:
  2023-02-09}.

\bibitem{nvidiaTuring}
Nvidia,
  \href{https://images.nvidia.com/aem-dam/en-zz/Solutions/design-visualization/technologies/turing-architecture/NVIDIA-Turing-Architecture-Whitepaper.pdf}{{NVIDIA
  Turing GPU Architecture}}, Tech. rep. (2018).
\newline\urlprefix\url{https://images.nvidia.com/aem-dam/en-zz/Solutions/design-visualization/technologies/turing-architecture/NVIDIA-Turing-Architecture-Whitepaper.pdf}

\bibitem{nvidiaAda}
Nvidia,
  \href{https://images.nvidia.com/aem-dam/Solutions/geforce/ada/nvidia-ada-gpu-architecture.pdf}{{NVIDIA
  Ada GPU Architecture}}, Tech. rep. (2022).
\newline\urlprefix\url{https://images.nvidia.com/aem-dam/Solutions/geforce/ada/nvidia-ada-gpu-architecture.pdf}

\bibitem{soman2010efficient}
J.~Soman, M.~K. Kumar, K.~Kothapalli, P.~Narayanan, Efficient discrete range
  searching primitives on the {GPU} with applications, in: 2010 International
  Conference on High Performance Computing, IEEE, 2010, pp. 1--10.

\bibitem{soman2012discrete}
J.~Soman, K.~Kothapalli, P.~Narayanan, Discrete range searching primitive for
  the {GPU} and its applications, Journal of Experimental Algorithmics (JEA) 17
  (2012) 4--1.

\bibitem{schieber1988finding}
B.~Schieber, U.~Vishkin, On finding lowest common ancestors: Simplification and
  parallelization, SIAM Journal on Computing 17~(6) (1988) 1253--1262.

\bibitem{salmon2019exploiting}
J.~Salmon, S.~McIntosh-Smith, Exploiting hardware-accelerated ray tracing for
  monte carlo particle transport with openmc, in: 2019 IEEE/ACM Performance
  Modeling, Benchmarking and Simulation of High Performance Computer Systems
  (PMBS), IEEE, 2019, pp. 19--29.

\bibitem{Wald2019RTXBR}
I.~Wald, W.~Usher, N.~Morrical, L.~M. Lediaev, V.~Pascucci, {RTX} beyond ray
  tracing: exploring the use of hardware ray tracing cores for tet-mesh point
  location, Proceedings of the Conference on High-Performance Graphics (2019).

\bibitem{Evangelou2021RadiusSearch}
I.~Evangelou, G.~Papaioannou, K.~Vardis, A.~A. Vasilakis, Fast radius search
  exploiting ray tracing frameworks, Journal of Computer Graphics Techniques
  (JCGT) 10~(1) (2021) 25--48.

\bibitem{Baxter:2016:M2}
S.~Baxter, moderngpu 2.0, \url{https://github.com/moderngpu/moderngpu/wiki}
  (2016).

\end{thebibliography}
\end{document}